\documentclass[aps,prd,onecolumn,notitlepage,superscriptaddress,longbibliography,nofootinbib]{revtex4-1}

\usepackage{dcolumn}    
\usepackage{bm}         
\usepackage{hyperref}   
\usepackage{bmpsize}
\usepackage{url}
\hypersetup{breaklinks=true}
\usepackage[english]{babel}
\usepackage[skip=1pt, justification=raggedright]{caption}
\usepackage{graphicx}
\vspace{-10pt}
\usepackage{amsmath}
\hypersetup{
    colorlinks=true,
    citecolor=blue,
    linkcolor=blue,
    urlcolor=blue
}

\begin{document}

\title{Probing Yukawa Gravity with Modulated Newtonian Cancellation in the Torsion bar type detector}

\author{Yuki Inoue}
\thanks{Corresponding author: iyuki@ncu.edu.tw}
\affiliation{Department of Physics,National Central University, Taoyuan, Taiwan}
\affiliation{Center for High Energy and High Field (CHiP), National Central University,  Taoyuan, Taiwan}
\affiliation{Institute of Physics, Academia Sinica, Taipei, Taiwan}
\affiliation{Institute of Particle and Nuclear Studies, High Energy Acceleration Research Organization (KEK), Tsukuba, Japan}

\author{Hsiang-Yu Huang}
\affiliation{Department of Physics,National Central University, Taoyuan, Taiwan}
\affiliation{Center for High Energy and High Field (CHiP), National Central University,  Taoyuan, Taiwan}

\author{Vivek Kumar}
\affiliation{Department of Physics,National Central University, Taoyuan, Taiwan}
\affiliation{Center for High Energy and High Field (CHiP), National Central University,  Taoyuan, Taiwan}

\author{Daiki Tanabe}
\affiliation{Center for High Energy and High Field (CHiP), National Central University,  Taoyuan, Taiwan}
\affiliation{Institute of Physics, Academia Sinica, Taipei, Taiwan}


\date{\today}

\begin{abstract}

We investigate the sensitivity of a torsion-bar gravitational-wave detector to Yukawa-type deviations from Newtonian gravity using a differential gravitational calibrator (GCal), where two rotating mass systems cancel the leading Newtonian torque.
We derive an exact expression for the residual torque and map the Yukawa signal into a strain-equivalent response in the sub-Hz band. We evaluate the sensitivity in the $(\alpha_Y,\lambda)$ parameter space, finding optimal performance at scales comparable to the experimental geometry, reaching $|\alpha_Y| = 2.4\times10^{-5}$ at $\lambda = 8,\mathrm{m}$.
The sensitivity is limited by residual Newtonian torque from imperfect cancellation rather than statistical noise, with a systematic floor reached at $T_{\rm eq} \simeq 9.25\times10^{4},\mathrm{s}$ ($\sim 26$ hours). This limit is dominated by uncertainties in the source-mass geometry.
The differential configuration retains sensitivity even at large interaction ranges, enabling constraints at meter-scale distances. These results establish torsion-bar detectors as a systematics-limited probe of non-Newtonian gravity in the sub-Hz band.

\end{abstract}

\maketitle
\section{Introduction}

Testing gravity at short distances and low frequencies provides a unique
window into physics beyond the Standard Model.
While Newtonian gravity has been verified over a wide range of length
scales, a variety of theoretical frameworks predict deviations from the
inverse-square law at sub-meter scales~\cite{Fischbach1999,Adelberger2003}.
Such deviations are often parameterized in terms of a Yukawa-type
correction to the gravitational potential~\cite{Fischbach1999},
characterized by a strength $(\alpha_Y)$ and a range $(\lambda)$.

Experimental searches for Yukawa interactions have been actively pursued
using torsion balances, atom interferometry, and precision force
measurements~\cite{Adelberger2009,Kapner2007,Geraci2008,Raffai2011_PRD},
placing strong constraints over a broad range of $(\alpha_Y,\lambda)$
parameter space.
More recently, constraints have also been derived from planetary motion,
satellite tracking, and astrophysical observations
\cite{Konopliv2011,Hees2017,Lee2017_GC}.

In parallel, gravitational-wave detector technologies have advanced
rapidly, opening new opportunities to probe weak forces with exquisite
sensitivity.
The direct detection of gravitational waves by LIGO~\cite{LIGO2016} and the
subsequent observations of compact binary mergers~\cite{GWTC1,GWTC2,GWTC3}
have established interferometric detectors as precision probes of
fundamental physics.
In particular, torsion-bar antennas, such as TOBA and TorPeDO operating in the sub-Hz band offer a
promising platform for detecting low-frequency gravitational signals ~\cite{Ando2010,TorPeDO2019}.
The CHRONOS detector, a cryogenic sub-Hz torsion-bar interferometer with
a quantum non-demolition speed-meter configuration~\cite{Inoue2025_CHRONOS,Inoue2025_CHRONOS_Optics,inoue2026chronosscienceprogram}, is also designed to
achieve high sensitivity in the $(0.1\text{--}10,\mathrm{Hz})$ band.
In this regime, the response of the torsion-bar system is dominated by
inertia, providing a clean and well-understood transfer function between
external torques and the measured strain-equivalent signal~\cite{Saulson1990}.

A key technique for absolute calibration in such systems is the
gravitational calibrator (GCal), in which rotating masses generate a
well-defined, time-dependent gravitational torque.
In existing gravitational-wave detectors such as LIGO, Virgo, and KAGRA,
calibration is primarily performed using photon calibrators (PCal),
which apply a known radiation-pressure force to the test masses~\cite{Karki2016,Goetz2010_LIGO_Calibration,Inoue2023_KAGRA_PCal}.
In parallel, Newtonian calibrators (NCal) generate a dynamic gravitational field using rotating masses and have been implemented as an independent calibration method~\cite{Estevez2018NCal,Acernese2018VirgoNCal}.
Early demonstrations of dynamic gravitational field generation and
detection date back several decades~\cite{Forward1967_DynamicGravity,Hirakawa1980,Ogawa1982,Kuroda1985,Astone1991,Astone1998},
and have recently been developed for modern interferometric detectors as
Newtonian calibrators~\cite{Estevez2018NCal}.
Unlike conventional displacement-based calibration methods such as photon
calibrators, the torque-coupled configuration directly excites the rotational
degree of freedom of the torsion bar, enabling efficient calibration in
the sub-Hz frequency band.
Moreover, by employing a differential configuration with two rotating
mass systems, it is possible to suppress the leading Newtonian
contribution through geometric cancellation~\cite{Inoue2018_GCal,inoue2026improvingcalibrationaccuracytorque}.

This cancellation scheme provides a powerful method for probing
non-Newtonian gravity.
Because a Yukawa interaction does not follow a pure power-law dependence
on distance, the cancellation condition for the Newtonian term does not
generally eliminate the Yukawa contribution.
As a result, a residual signal remains that is directly sensitive to the
parameters $(\alpha_Y,\lambda)$.
This approach allows one to transform a calibration system into a probe
of short-range gravitational physics.

In this paper, we develop a framework for evaluating the sensitivity of
a torsion-bar detector to Yukawa-type deviations from Newtonian gravity
using a differential GCal configuration.
We derive the exact expression for the residual torque after imposing
the Newtonian cancellation condition, without relying on low-order
approximations.
The resulting torque is then converted into a strain-equivalent signal
using the mechanical transfer function of the torsion-bar system.
Based on this formulation, we compute the signal-to-noise ratio and
project constraints on the Yukawa parameters $(\alpha_Y,\lambda)$.

An important aspect of this method is the role of systematic
uncertainties.
In particular, imperfect cancellation of the Newtonian contribution
introduces a residual background torque that can limit the achievable
sensitivity.
This issue is closely related to calibration accuracy requirements in
gravitational-wave detectors~\cite{Hall2017CalibrationRequirements,Cahillane2017_Calibration}.
We show that this residual sets a fundamental floor on the constraint on
$(\alpha_Y)$, independent of the observation time, and must be carefully
controlled in the experimental design.

This paper is organized as follows.
In Sec.~II, we derive the exact expression for the Yukawa-induced torque
in the differential GCal configuration.
In Sec.~III, we convert the torque into a strain-equivalent signal using
the mechanical response of the torsion bar.
In Sec.~IV, we evaluate the sensitivity and present projected constraints
in the $(\alpha_Y,\lambda)$ parameter space, including the impact of
systematic uncertainties.

\section{Torque Expansion with Yukawa Correction}
\label{sec:yukawa_torque_derivation}

Before presenting the detailed derivation, we first summarize the core concept of the experimental configuration.

\begin{figure}[t]
\centering
\includegraphics[width=0.8\linewidth]{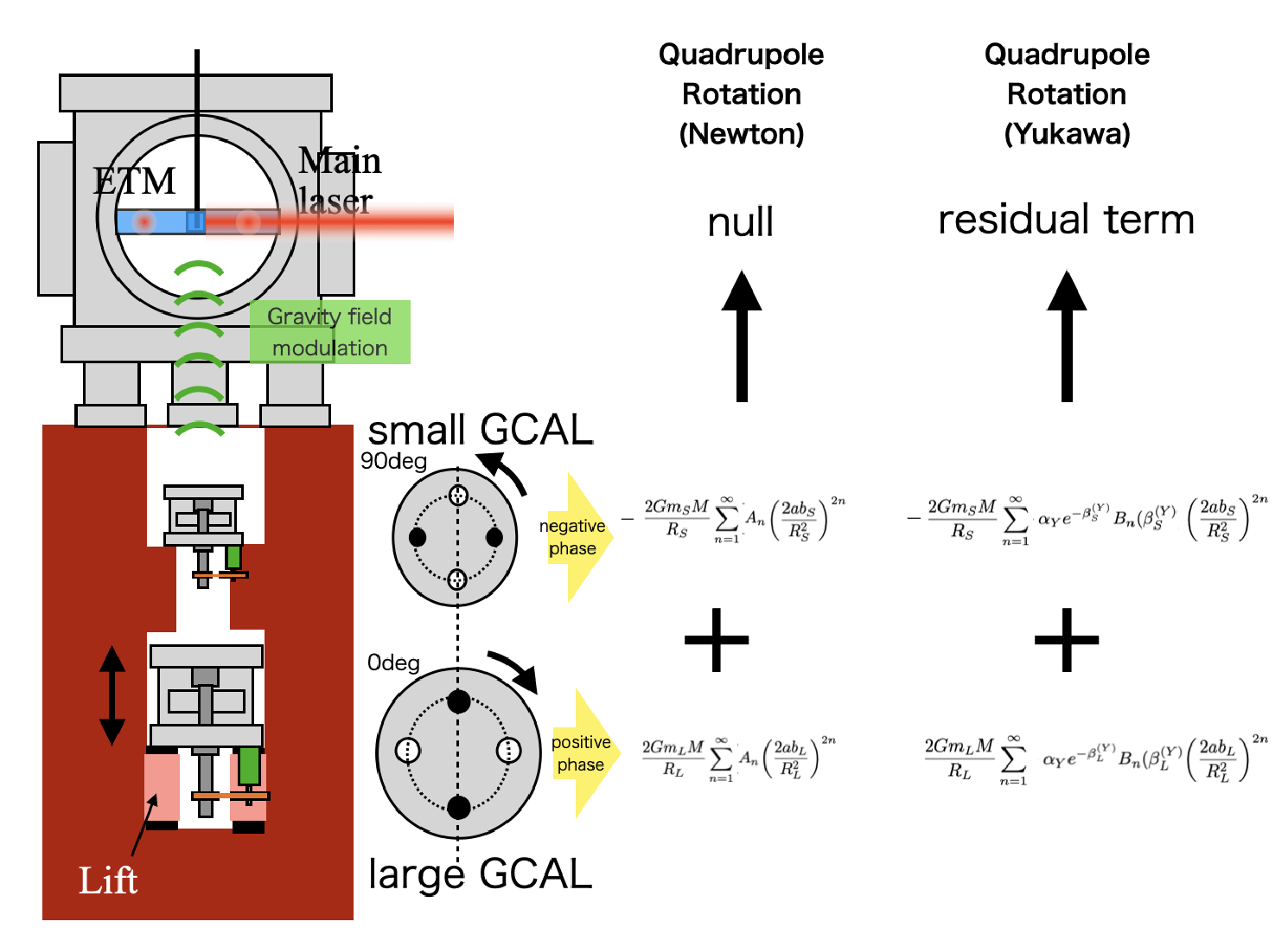}
\caption{
Schematic diagram of the differential GCal configuration.
Two source masses are placed under the torsion bar with a controlled geometric scaling.
The configuration is designed such that the Newtonian torque cancels,
while a residual Yukawa contribution remains. Large GCal can be tuned by lift to optimize the position.
}
\label{fig:concept}
\end{figure}

As shown in Fig.~\ref{fig:concept}, the experiment employs a differential configuration
in which two gravitational field calibrators (GCals) are arranged symmetrically
with respect to the torsion bar.
By appropriately choosing the geometric scaling between the two GCals,
the Newtonian contributions from each side cancel to a high degree.
Such dynamic gravitational field configurations have been studied since early experimental works~\cite{Forward1967_DynamicGravity,Hirakawa1980}
and form the basis of modern Newtonian calibrators~\cite{Estevez2018NCal,Acernese2018VirgoNCal}
as well as the torque-coupled GCal configuration developed in Ref.~\cite{Inoue2018_GCal,inoue2026improvingcalibrationaccuracytorque}.

This cancellation relies on the fact that the Newtonian interaction follows a pure inverse-distance law,
allowing the torque contributions to be balanced by tuning the geometry.
In contrast, a Yukawa-type interaction introduces an additional exponential factor,
which depends on the absolute distance scale.
As a result, the cancellation condition for the Newtonian term does not apply to the Yukawa contribution.

Consequently, after the cancellation of the dominant Newtonian torque,
a residual signal remains that is sensitive to the Yukawa interaction.
The experiment therefore converts a large common-mode Newtonian signal
into a small differential signal that directly probes deviations from Newtonian gravity.

With this conceptual picture in mind, we now proceed to derive the Yukawa-corrected torque.

We start from the modified gravitational potential~\cite{Raffai2011_PRD}
\begin{equation}
V(r)
=
-\frac{GmM}{r}\left(1+\alpha_Y e^{-r/\lambda}\right),
\label{eq:yukawa_potential}
\end{equation}
where $G$ is the gravitational constant, $m$ and $M$ are the source and test masses,
$\alpha_Y$ is the Yukawa coupling strength, and $\lambda$ is the interaction range.

The potential can be separated into the Newtonian and Yukawa parts as
\begin{equation}
V(r)=V_{\mathrm N}(r)+V_{\mathrm Y}(r),
\end{equation}
with
\begin{align}
V_{\mathrm N}(r) &= -\frac{GmM}{r},\
V_{\mathrm Y}(r) &= -\alpha_Y \frac{GmM}{r}e^{-r/\lambda}.
\end{align}

Since the Newtonian contribution has already been derived in previous work~\cite{inoue2026improvingcalibrationaccuracytorque}, we focus on the Yukawa contribution in the following.

\subsection{Geometrical parameterization}

\begin{figure}[t]
\centering
\includegraphics[width=0.8\linewidth]{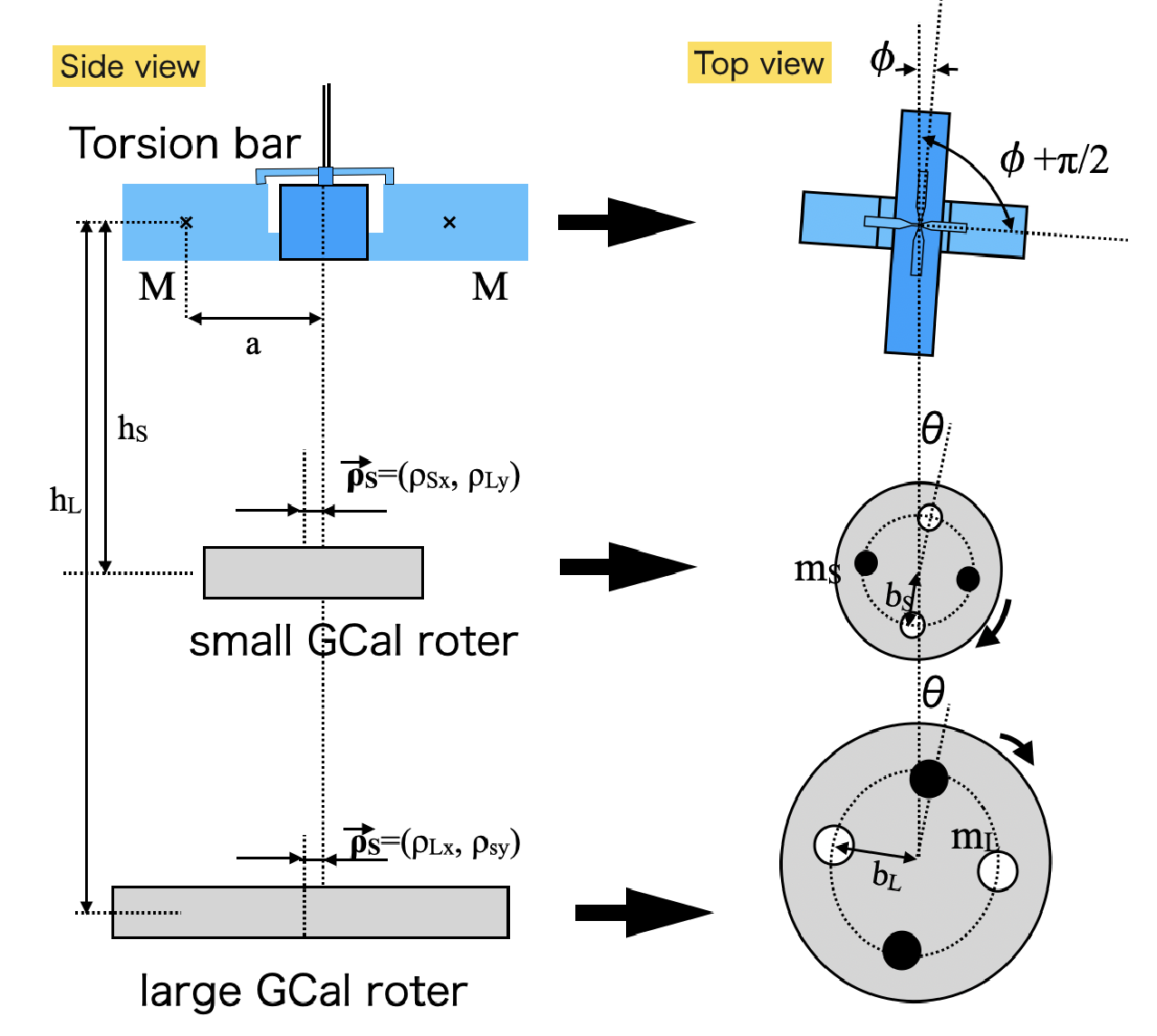}
\caption{
Geometrical configuration used for the torque calculation.
The parameters $a_i$, $b_i$, $\rho_i$ and $R_i$ define the relative positions
of the rotor masses and the test mass, where $i={S,L}$.
The relative angle between the rotor and the torsion bar
is denoted by $\Delta = \theta - \phi$.
}
\label{fig:geometry}
\end{figure}

The geometrical configuration used in the torque calculation is shown in Fig.~\ref{fig:geometry}.

We introduce a reference distance scale
\begin{equation}
R^2 = a^2 + b^2 + h^2 + \rho_\perp^2,
\end{equation}
which represents the typical separation between the source masses and the test mass.
The actual distance can then be written as a modulation around this reference scale,
\begin{equation}
r^2 = R^2 + \delta,
\end{equation}
where $\delta$ encodes the angular dependence of the geometry.

The modulation $\delta$ arises from the relative orientation
between the rotor and the torsion bar.
Defining the rotor phase $\theta$ and the bar orientation $\phi$,
the relative angle is
\begin{equation}
\Delta = \theta - \phi.
\end{equation}
The geometric coupling between the two directions appears through $\cos\Delta$.

It is convenient to introduce a dimensionless parameter
\begin{equation}
u \equiv -\frac{\delta}{R^2},
\end{equation}
so that the distance takes the form
\begin{equation}
r = R \sqrt{1 - u}.
\label{eq:r_param}
\end{equation}

The parameter $u$ can be decomposed into contributions associated with different geometric effects.
In particular, the dominant term arises from the relative-angle modulation,
\begin{equation}
u \simeq \frac{2ab}{R^2}\cos\Delta.
\label{eq:u_dominated}
\end{equation}

This expression shows that the distance modulation is primarily governed by
the relative phase $\Delta$ between the rotor and the torsion bar,
as shown in Fig.~\ref{fig:geometry}.
Other contributions, such as small offsets of the source masses,
provide subleading corrections and are neglected in the following.

Using this parametrization, the Yukawa part of the potential can be written as
\begin{equation}
V_{\mathrm Y}
=
-\alpha_Y \frac{GmM}{R}
\frac{e^{-\beta^{(Y)} \sqrt{1-u}}}{\sqrt{1-u}},
\qquad
\beta^{(Y)} \equiv \frac{R}{\lambda}.
\label{eq:yukawa_kernel_def}
\end{equation}

The problem is therefore reduced to expanding the kernel
\begin{equation}
K(u;\beta^{(Y)})
\equiv
\frac{e^{-\beta^{(Y)} \sqrt{1-u}}}{\sqrt{1-u}}
\label{eq:kernel_definition}
\end{equation}
in powers of $u$.
\subsection{Expansion of the Yukawa kernel}

We first rewrite the kernel as
\begin{equation}
K(u;\beta^{(Y)})
=
e^{-\beta^{(Y)}}
\frac{\exp\!\left[\beta^{(Y)}\left(1-\sqrt{1-u}\right)\right]}{\sqrt{1-u}}.
\label{eq:kernel_rewrite}
\end{equation}
This form is convenient because the factor $e^{-\beta^{(Y)}}$ can be taken out, while the remaining expression is analytic around $u=0$.

We now expand $K(u;\beta^{(Y)})$ in a Taylor series,
\begin{equation}
K(u;\beta^{(Y)})
=
e^{-\beta^{(Y)}}
\sum_{k=0}^{\infty} c_k(\beta^{(Y)})\, u^k.
\label{eq:kernel_taylor}
\end{equation}
The coefficients $c_k(\beta^{(Y)})$ are known to be expressible in terms of the Bessel polynomials $y_k(\beta^{(Y)})$:
\begin{equation}
c_k(\beta^{(Y)})=\frac{y_k(\beta^{(Y)})}{2^k k!}.
\label{eq:ck_bessel_poly}
\end{equation}

Hence the Yukawa potential becomes
\begin{equation}
V_{\mathrm Y}
=
-\alpha_Y \frac{GmM}{R} e^{-\beta^{(Y)}}
\sum_{k=0}^{\infty}
\frac{y_k(\beta^{(Y)})}{2^k k!}u^k.
\label{eq:yukawa_potential_series}
\end{equation}

\subsection{Torque from the Yukawa potential for 2f components}

As in the Newtonian case, the variable $u$ encodes the angular dependence of the rotor--test-mass configuration.
We define the phase difference as
\begin{equation}
\Delta = \phi + \theta,
\end{equation}
where $\phi$ is the rotation angle of the rotor and $\theta$ is a constant geometrical offset.
Since $\theta$ is constant, one has
\begin{equation}
\frac{\partial}{\partial \phi} = \frac{\partial}{\partial \Delta}.
\end{equation}
In the present convention, we set $\theta=0$, so that $\Delta=\phi$.

For the symmetric quadrupolar geometry considered here, the angular dependence is such that only even powers of the basic geometrical quantity contribute to the leading torque term.
After summing over the symmetric mass configuration and differentiating with respect to $\Delta$, the leading harmonic structure is proportional to $\sin(2\Delta)$.
More specifically, the geometrical part of the derivation proceeds exactly as in the Newtonian case, and one finds that only the terms $u^{2n}$ survive in the final quadrupole torque series.
The Yukawa correction therefore preserves the same angular structure as the Newtonian torque, while modifying the coefficient of each multipole order.

The torque is given by
\begin{equation}
\tau_\phi^{(\mathrm Y)}
=
- \frac{\partial V_{\mathrm Y}}{\partial \phi}
=
- \frac{\partial V_{\mathrm Y}}{\partial \Delta}.
\label{eq:torque_def}
\end{equation}

Substituting Eq.~\eqref{eq:yukawa_potential_series}, we obtain
\begin{equation}
\tau_\phi^{(\mathrm Y)}
=
\alpha_Y \frac{GmM}{R} e^{-\beta^{(Y)}}
\sum_{k=0}^{\infty}
\frac{y_k(\beta^{(Y)})}{2^k k!}
\frac{\partial}{\partial \Delta} u^k.
\label{eq:torque_before_sym}
\end{equation}

At this stage, the same geometrical manipulations as in the Newtonian derivation can be used.
The result is that the torque can be written in the form
\begin{equation}
\tau_\phi^{(\mathrm Y)}
=
\frac{2GmM}{R}
\alpha_Y e^{-\beta^{(Y)}}
\sum_{n=1}^{\infty}
B_n(\beta^{(Y)})
\left(\frac{2ab}{R^2}\right)^{2n}
\sin(2\Delta),
\label{eq:yukawa_torque_final}
\end{equation}
where the Yukawa coefficient $B_n(\beta^{(Y)})$ is given by
\begin{equation}
B_n(\beta^{(Y)})
=
\frac{y_{2n}(\beta^{(Y)})}{2^{2n}(2n)!}
\frac{\binom{2n-1}{n-1}}{2^{2n-1}}.
\label{eq:Bn_def}
\end{equation}

Using this formulation, the derivation of the Newtonian contribution is summarized in Appendix~\ref{app:newton}.
The general Newtonian coefficient $A_n$ has been derived in previous work~\cite{inoue2026improvingcalibrationaccuracytorque} and is given in Eq.~\eqref{eq:An_def}.
Importantly, we show that this coefficient can be consistently recovered as the $\lambda \to \infty$ limit of the Yukawa expression, i.e.,
\begin{equation}
\lim_{\beta^{(Y)} \to 0} B_n(\beta^{(Y)}) = A_n,
\label{eq:lim_Bn_An}
\end{equation}
demonstrating that Eq.~\eqref{eq:yukawa_torque_final} provides a unified generalization of the Newtonian expansion.

Including both the Newtonian and Yukawa contributions, the total torque is written as
\begin{equation}
\small
\tau
=
\frac{2GmM}{R}
\sum_{n=1}^{\infty}
\left[
A_n
+
\alpha_Y e^{-\beta^{(Y)}} B_n(\beta^{(Y)})
\right]
\left(\frac{2ab}{R^2}\right)^{2n}
\sin(2\Delta).
\label{eq:total_torque_with_yukawa}
\end{equation}

Thus, the Yukawa correction introduces a scale-dependent modification through the factor $e^{-R/\lambda}$ and the polynomial dependence encoded in the Bessel polynomials $y_{2n}(R/\lambda)$, while preserving the same quadrupolar angular dependence $\sin(2\Delta)$ as in the Newtonian case.

\section{Newtonian cancellation}

We consider a differential GCal configuration in which two GCals are
placed on opposite sides of the torsion bar along the vertical direction
and driven with opposite rotation senses.
With this configuration, the leading Newtonian contribution can be
suppressed, while a residual Yukawa contribution remains due to its
non-power-law distance dependence.
This provides a direct method to probe short-range deviations from
Newtonian gravity.
Such differential configurations have been proposed in the context of
gravity-field-based calibration and tests of non-Newtonian gravity~\cite{Inoue2018_GCal,Raffai2011_PRD}.

We denote the GCal located farther from the torsion bar by the subscript
$L$ (`long'') and the one located closer to the torsion bar by $S$
(`short'').
The torque generated by each GCal is given by
\begin{align}
\tau_{L}
&=
\frac{2Gm_{L}M}{R_{L}}
\sum_{n}
\Bigl[A_n+\alpha_Y e^{-\beta^{(Y)}*{L}} B_n\Bigr]
\left(\frac{2ab*{L}}{R_L^2}\right)^{2n}
\sin(2 \omega t),
\\
\tau_{S}
&=
\frac{2Gm_{S}M}{R_{S}}
\sum_{n}
\Bigl[A_n+\alpha_Y e^{-\beta^{(Y)}*{S}} B_n\Bigr]
\left(\frac{2ab*{S}}{R_S^2}\right)^{2n}
\sin(-2 \omega t),
\end{align}
where $\beta^{(Y)}_L \equiv R_L/\lambda$ and $\beta^{(Y)}_S \equiv R_S/\lambda$.

Using $\sin(-2\omega t)=-\sin(2\omega t)$, the total torque becomes
\begin{align}
\tau_{L}+\tau_{S}
&= 2GM
\sum_{n=1}^{\infty}
\Bigl[
A_n,\mathcal{N}_n
+
\alpha_Y,\mathcal{Y}_n
\Bigr]
\sin(2\omega t),
\label{eq:tau_sum_compact}
\end{align}
where
\begin{align}
\mathcal{N}_n
&\equiv
\frac{m_L}{R_L}
\left(\frac{2ab_L}{R_L^2}\right)^{2n}
-
\frac{m_S}{R_S}
\left(\frac{2ab_S}{R_S^2}\right)^{2n},
\\
\mathcal{Y}_n
&\equiv
e^{-\beta^{(Y)}_L} B_n(\beta^{(Y)}_L)
\frac{m_L}{R_L}
\left(\frac{2ab_L}{R_L^2}\right)^{2n}
\nonumber\\
&\quad -
e^{-\beta^{(Y)}_S} B_n(\beta^{(Y)}_S)
\frac{m_S}{R_S}
\left(\frac{2ab_S}{R_S^2}\right)^{2n}.
\end{align}

The crucial point is that the Newtonian cancellation condition must be imposed
on the \emph{fully resummed} Newtonian torque, rather than order by order in the multipole expansion.
This distinction is essential because the series does not truncate,
and cancellation at a finite order does not guarantee suppression of the total Newtonian contribution.

Using the generating function
\begin{equation}
\sum_{n=0}^{\infty} A_n x^n = (1-x)^{-1/2},
\qquad
A_n=\frac{(2n)!}{(n!)^2 4^n},
\end{equation}
the Newtonian part of the torque generated by a single GCal can be
summed exactly as
\begin{equation}
\tau_N(R,b)
=\frac{2GmM}{R}
\left[
\frac{1}{\sqrt{1-\dfrac{4a^2b^2}{R^4}}}-1
\right]
\sin(2\omega t).
\label{eq:tau_newton_resummed}
\end{equation}

Therefore, for the differential configuration, the total Newtonian torque is
\begin{align}
\tau_{N,L}+\tau_{N,S}
=2GM \Bigg[
&\frac{m_L}{R_L}
\left(
\frac{1}{\sqrt{1-\dfrac{4a^2b_L^2}{R_L^4}}}-1
\right)
\nonumber\\
&-
\frac{m_S}{R_S}
\left(
\frac{1}{\sqrt{1-\dfrac{4a^2b_S^2}{R_S^4}}}-1
\right)
\Bigg]\sin(2\omega t).
\label{eq:tau_newton_total_resummed}
\end{align}

Hence, the exact Newtonian null condition is given by
\begin{equation}
\frac{m_L}{R_L}
\left[
\frac{1}{\sqrt{1-\dfrac{4a^2b_L^2}{R_L^4}}}-1
\right]
=\frac{m_S}{R_S}
\left[
\frac{1}{\sqrt{1-\dfrac{4a^2b_S^2}{R_S^4}}}-1
\right].
\label{eq:newton_null_exact}
\end{equation}

Since
\begin{equation}
R_L^2=a^2+b_L^2+h_L^2+\rho_L^2,
\qquad
R_S^2=a^2+b_S^2+h_S^2+\rho_S^2,
\end{equation}
this condition determines the required long-side distance $h_L$
at which the \emph{fully resummed} Newtonian contribution vanishes.

Thus, the Newtonian cancellation is achieved not by truncating the
multipole expansion at a finite order, but by tuning the geometry
such that the exact Newtonian torque becomes zero.

In the practically relevant case
\begin{equation}
m \equiv m_L = m_S,
\qquad
2b \equiv b_L = 2b_S ,
\end{equation}
the null condition reduces to a single equation relating $h_L$ and $h_S$.
Although this condition can be written analytically, it does not in
general yield a simple closed-form expression for $h_L$.
Therefore, for a fixed short-side geometry, we determine the long-side
distance $h_L$ numerically by solving Eq.~(\ref{eq:newton_null_exact}).

It is then convenient to parametrize the null condition in the form
\begin{equation}
\gamma h \equiv h_L = \gamma h_S,
\label{eq:gamma_def_h}
\end{equation}
where $\gamma$ is the numerically determined ratio that realizes the
exact Newtonian cancellation for the chosen geometry.

Thus, the Newtonian-null configuration is fully specified by the root of
Eq.~(\ref{eq:newton_null_exact}), or equivalently by the corresponding value of
$\gamma$.

Having fixed the geometry such that the exact Newtonian contribution
vanishes, we now evaluate the remaining Yukawa component without
truncating the multipole expansion.

Once this condition is imposed, the Newtonian contribution
vanishes identically, and the residual signal is purely Yukawa:
\begin{equation}
{\footnotesize
\tau_Y(\lambda)
=2GM,\alpha_Y
\sum_{n=1}^{\infty}
\mathcal{Y}_n(\lambda)\sin(2\omega t)
}
\label{eq:tau_yukawa_exact_series}
\end{equation}

For compactness, we define
$
{R}_S \equiv R_S(a,b,\rho,h,\gamma),;
{R}_L \equiv R_L(a,b,\rho,h,\gamma)
$,
and express all quantities in terms of these functions.

Then,
\begin{equation}
\footnotesize
\begin{aligned}
\mathcal{Y}_n(\lambda)
&=
\frac{m}{\mathcal{R}_S}
\left(\frac{b}{\mathcal{R}_S}\right)^{2n}
\Bigl[
2^{2n}
\left(\frac{\mathcal{R}_S}{\mathcal{R}_L}\right)^{2n+1}
e^{-\mathcal{R}_L/\lambda}
B_n\left(\frac{\mathcal{R}_L}{\lambda}\right)
\\
&\qquad -
e^{-\mathcal{R}_S/\lambda}
B_n\left(\frac{\mathcal{R}_S}{\lambda}\right)
\Bigr]
\end{aligned}
\label{eq:Y_n_lambda}
\end{equation}

Therefore, the exact residual Yukawa torque at the Newtonian null is
\begin{equation}
\footnotesize
\begin{aligned}
\tau_Y(\lambda)
&=
2GM\alpha_Y
\frac{m}{\mathcal{R}_S}
\sum_{n=1}^{\infty}
\left(\frac{2ab}{\mathcal{R}_S^2}\right)^{2n}
\\
&\quad \times
\Bigl[
2^{2n}
\left(\frac{\mathcal{R}_S}{\mathcal{R}_L}\right)^{2n+1}
e^{-\mathcal{R}_L/\lambda}
B_n\left(\frac{\mathcal{R}_L}{\lambda}\right)
\\
&\qquad -
e^{-\mathcal{R}_S/\lambda}
B_n\left(\frac{\mathcal{R}_S}{\lambda}\right)
\Bigr]
\sin(2\omega t)
\end{aligned}
\label{eq:tau_yukawa_lambda}
\end{equation}

Here, $\mathcal{R}_S$ and $\mathcal{R}_L$ denote the exact geometric
quantities after imposing the parameter relations, so that the residual
torque is written explicitly as a function of the Yukawa range
$\lambda$.
The conversion from the torque signal to the strain-equivalent signal
$h(t)$ is summarized in Appendix~\ref{app:transfer_function}.

We next define the strain-equivalent signal generated by a possible
Yukawa-type deviation from Newtonian gravity.
Let the Yukawa torque amplitude at angular frequency
$\Omega = 2\omega$ be denoted by $\tau_Y(\Omega; \alpha_Y, \lambda)$,
where $\alpha_Y$ is the Yukawa strength and $\lambda$ is the Yukawa
interaction range.

The induced angular response of the torsion bar is given by
\begin{equation}
\theta_Y(\Omega; \alpha_Y, \lambda)
=
\chi(\Omega)\,\tau_Y(\Omega; \alpha_Y, \lambda),
\label{eq:theta_yukawa_def}
\end{equation}
where \(\chi(\Omega)\) is the mechanical susceptibility of the torsion-bar
degree of freedom.

Using the same strain-equivalent definition from Appendix~\ref{app:transfer_function}, the
Yukawa-induced strain signal is defined as
\begin{equation}
h_{\rm Yukawa}(\Omega; \alpha_Y, \lambda)
=
\left|
\frac{2\,\chi(\Omega)\,\tau_Y(\Omega; \alpha_Y, \lambda)}
{\eta_g \, |F_{\rm eff}|}
\right|.
\label{eq:h_yukawa_def}
\end{equation}
This quantity represents the apparent detector strain corresponding to a
Yukawa torque signal at frequency \(\Omega\).

In the high-frequency regime relevant for CHRONOS,
\(\Omega \gg \omega_0\), the mechanical susceptibility reduces to
\begin{equation}
\chi(\Omega)\simeq -\frac{1}{I\Omega^2},
\label{eq:chi_highf_repeat}
\end{equation}
where the typical torsional resonance frequency \(\omega_0\) is smaller than
\(0.01\,\mathrm{Hz}\), which is sufficiently below the operational frequency
range considered in this study.

Substituting this into Eq.~(\ref{eq:h_yukawa_def}), we obtain
\begin{equation}
h_{\rm Yukawa}(\Omega; \alpha_Y, \lambda)
\simeq
\frac{2\,|\tau_Y(\Omega; \alpha_Y, \lambda)|}
{\eta_g\,|F_{\rm eff}|\,I\,\Omega^2}.
\label{eq:h_yukawa_highf}
\end{equation}

Therefore, once the exact Yukawa torque
\(\tau_Y(\Omega; \alpha_Y, \lambda)\) is evaluated, the corresponding
strain-equivalent signal follows directly through the same transfer
function as for the standard GCal signal.
This formulation is particularly convenient for the sensitivity analysis
in the next section, where constraints on \((\alpha_Y, \lambda)\) can be
obtained by directly comparing \(h_{\rm Yukawa}\) with the detector strain
sensitivity.

\section{Systematic error limit and sensitivity}

Table~\ref{tab:prms} summarizes the fiducial parameter set used in this analysis,
together with their associated absolute uncertainties.
These parameters define both the gravitational field generated by the GCal
and the response of the torsion-bar detector, and therefore determine
the level of residual Newtonian contamination after cancellation.

\begin{table}[t]
\centering
\begin{tabular}{ccc}
\hline\hline
Parameters & Values & Absolute uncertainty \\
\hline
\multicolumn{3}{l}{\textit{Fundamental constant}} \\
\hline
$G$ & $6.67 \times 10^{-11}$ & $1.50 \times 10^{-15}$ \\
\hline
\multicolumn{3}{l}{\textit{GCal (source masses)}} \\
\hline
$m_S$ & $1.8842$ & $9.04 \times 10^{-5}$ \\
$m_L$ & $1.8842$ & $9.04 \times 10^{-5}$ \\
$h_S$ & $8.000$ & $1.00 \times 10^{-3}$ \\
$h_L=\gamma h_S$ & $10.553$ & $1.00 \times 10^{-3}$ \\
$b_S$ & $0.250$ & $2.50 \times 10^{-5}$ \\
$b_L$ & $0.500$ & $2.50 \times 10^{-5}$ \\
$\rho_S$ & $0.0010$ & $1.00 \times 10^{-3}$ \\
$\rho_L$ & $0.0010$ & $1.00 \times 10^{-3}$ \\
\hline
\multicolumn{3}{l}{\textit{Torsion-bar detector}} \\
\hline
$M$ & $85.50$ & $1.41 \times 10^{-4}$ \\
$a$ & $0.302$ & $2.50 \times 10^{-5}$ \\
\hline\hline
\end{tabular}
\caption{
The parameter set used in the sensitivity analysis.
All uncertainties are given as absolute values.
The long-baseline distance $h_L$ is determined by the Newtonian cancellation condition $h_L = \gamma h_S$.
}
\label{tab:prms}
\end{table}

In this section, we quantify how the parameter uncertainties listed in
Table~\ref{tab:prms} propagate into the residual Newtonian signal
after differential cancellation, and determine the corresponding
systematic error floor.
Rather than repeating the definition of the strain-equivalent signal
$h_N$, we focus on its uncertainty $\sigma_{h_N}$ and its decomposition.

To identify the dominant sources of uncertainty, we decompose the total
variance into contributions from individual parameters.
The fractional contribution of each parameter is defined as
\begin{equation}
C_i
=
\frac{\sigma_{h_N,i}^2}{\sum_j \sigma_{h_N,j}^2}\times 100~\%,
\label{eq:Ci_def}
\end{equation}
where $\sigma_{h_N,i}$ is the standard deviation induced by varying only
the parameter $p_i$ while keeping others fixed.
This decomposition is evaluated using a Monte Carlo propagation,
which captures the full nonlinear dependence of the Newtonian
cancellation condition.

\begin{table}[t]
\centering
\begin{tabular}{ccc}
\hline\hline
Parameter & $\sigma_{h_N}$,[-] & Contribution (\%) \\
\hline
$h_S$ & $4.63 \times 10^{-18}$ & 58.3 \\
$h_L$ & $3.51 \times 10^{-18}$ & 33.6 \\
$b_S$ & $1.47 \times 10^{-18}$ & 5.9 \\
$b_L$ & $7.38 \times 10^{-19}$ & 1.5 \\
$m_L$ & $3.58 \times 10^{-19}$ & 0.3 \\
$m_S$ & $3.57 \times 10^{-19}$ & 0.3 \\
$a$ & $1.86 \times 10^{-21}$ & 0.0 \\
$\rho_S$ & $7.13 \times 10^{-22}$ & 0.0 \\
$\rho_L$ & $4.10 \times 10^{-22}$ & 0.0 \\
$G$ & $1.34 \times 10^{-31}$ & 0.0 \\
$M$ & $2.15 \times 10^{-42}$ & 0.0 \\
\hline
\end{tabular}
\caption{
Noise budget of the residual Newtonian contribution expressed in strain.
The uncertainty $\sigma_{h_N}$ is dimensionless.
}
\label{tab:noise_budget}
\end{table}

As shown in Table~\ref{tab:noise_budget}, the residual uncertainty is
overwhelmingly dominated by the longitudinal geometric parameters
$h_S$ and $h_L$, which together account for more than $90\%$ of the
total variance.
The source-size parameters $b_S$ and $b_L$ contribute only at the
few-percent level, while all other parameters are negligible.

This hierarchy directly reflects the structure of the Newtonian
cancellation condition,
\begin{equation}
h_L = \gamma h_S,
\label{eq:hL_gamma_hS}
\end{equation}
which suppresses the leading-order Newtonian contribution.
Because the cancellation is achieved by balancing the longitudinal
geometry, small perturbations in $h_S$ or $h_L$ shift the cancellation
point and generate a first-order residual in $h_N$.
In contrast, variations in $b_S$ and $b_L$ enter only through higher-order
geometric dependence and therefore have a subdominant impact on the
systematic floor.
The residual systematic uncertainty defines the integration time beyond
which further averaging no longer improves the measurement.
We therefore introduce the equivalent integration time $T_{\rm eq}$ as
the observation time at which the statistical strain sensitivity reaches
the residual Newtonian systematic floor.

\subsection{Dependence on source geometry}

To study the geometric dependence of the systematic floor, we vary the
source radius $b_L$ and, for each configuration, determine the
corresponding cancellation parameter $\gamma$, the residual Newtonian
uncertainty $\sigma_{h_N}$, and the equivalent integration time
$T_{\rm eq}$.

\begin{table}[t]
\centering
\caption{
Dependence of the residual Newtonian uncertainty on the source radius $b_L$.
The equivalent integration time $T_{\rm eq}$ is evaluated for each configuration.
}
\begin{tabular}{cccc}
\hline\hline
$b_L$ (m) & $\gamma$ & $\sigma_{h_N}$ & $T_{\rm eq}$ (sec) \\
\hline
0.10 & 1.3199 & $4.06 \times 10^{-19}$ & $2.05 \times 10^{7}$ \\
0.20 & 1.3198 & $1.14 \times 10^{-18}$ & $2.58 \times 10^{6}$ \\
0.30 & 1.3196 & $2.32 \times 10^{-18}$ & $6.27 \times 10^{5}$ \\
0.40 & 1.3194 & $3.95 \times 10^{-18}$ & $2.16 \times 10^{5}$ \\
0.50 & 1.3191 & $6.04 \times 10^{-18}$ & $9.25 \times 10^{4}$ \\
\hline
\end{tabular}
\label{tab:bscan}
\end{table}

Table~\ref{tab:bscan} shows that the residual Newtonian uncertainty
increases monotonically with $b_L$, while the equivalent integration
time decreases accordingly.
Thus, a larger source radius raises the systematic floor and reduces the
amount of statistically useful integration time.

At the same time, the scaling parameter $\gamma$ remains nearly constant
($\gamma \simeq 1.319$) over the explored range, indicating that the
cancellation condition is controlled primarily by the longitudinal
geometry and only weakly modified by the transverse source size.
Here, $\gamma$ is obtained by solving the exact Newtonian cancellation
condition given in Eq.~\eqref{eq:newton_null_exact}.

This behavior implies a nontrivial design trade-off.
A larger $b_L$ generally enhances the Yukawa signal amplitude, but it
also increases the residual Newtonian background.
Conversely, a smaller $b_L$ lowers the systematic floor and allows a
longer effective integration time before the measurement becomes
systematics limited.
The final sensitivity must therefore be evaluated by consistently taking
both effects into account.

In this analysis, we assume tungsten as the source mass material.
The detailed mechanical and gravitational response of tungsten is
discussed in Inoue et al~\cite{Inoue2025_CHRONOS,Inoue2025_CHRONOS_Optics,inoue2026chronosscienceprogram}.

\subsection{Sensitivity curve}

The projected constraint is obtained from
\begin{equation}
\alpha_{Y{\rm lim}}(\lambda)
=
\frac{2\sqrt{S_h(\Omega)}}
{h_{\rm Yukawa}(\Omega;1,\lambda)\sqrt{T_{\rm eq}}},
\label{eq:alpha_lim}
\end{equation}
where $T_{\rm eq}$ is the equivalent integration time defined by the
condition that the accumulated statistical sensitivity reaches the
residual Newtonian systematic floor.
Thus, the final sensitivity is evaluated not for an arbitrarily chosen
observation time, but for the maximum effective integration time allowed
by the Newtonian systematic uncertainty for each configuration.

\begin{figure}[t]
\centering
\includegraphics[width=0.9\linewidth]{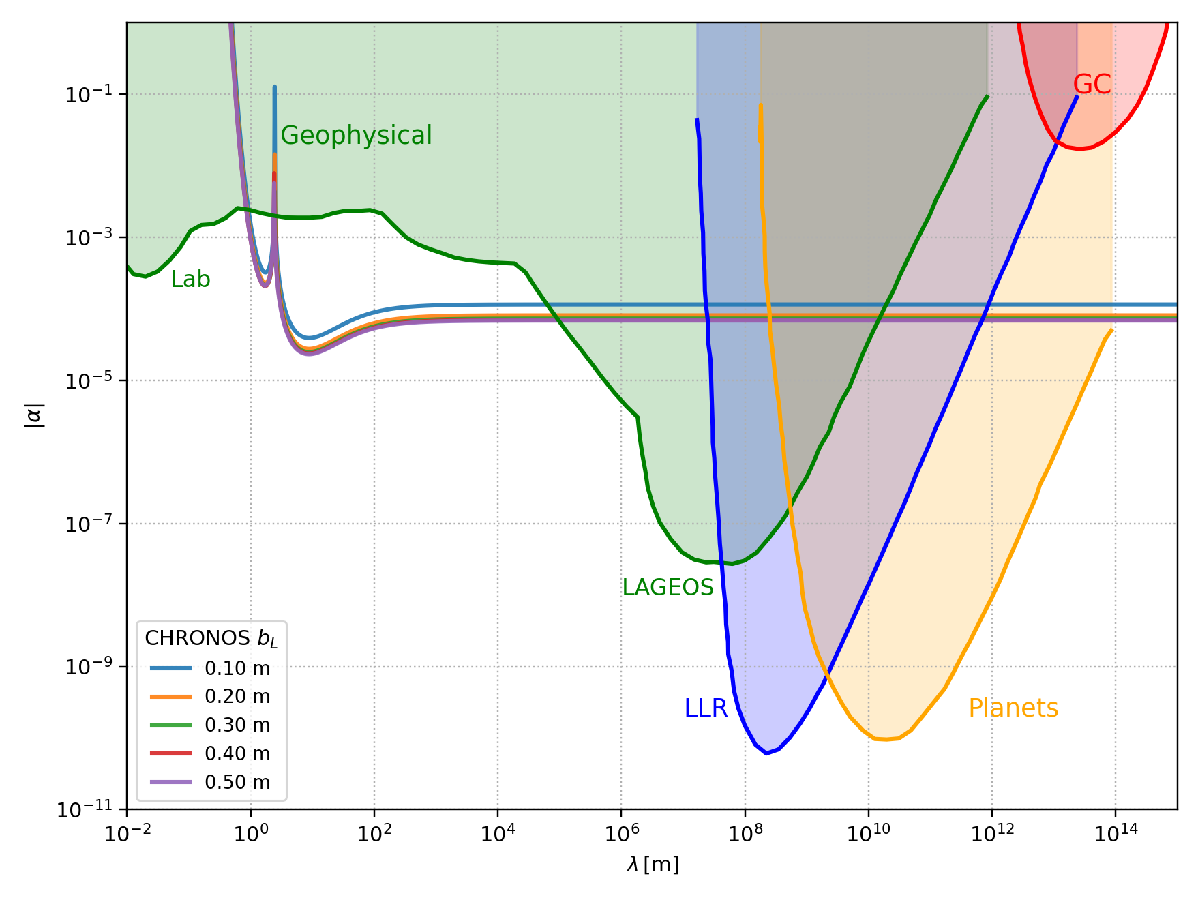}
\caption{
Projected sensitivity to the Yukawa coupling strength $\alpha_Y$ as a function of the interaction range $\lambda$.
Different curves correspond to different values of $b_L$.
Existing experimental and astrophysical constraints (Lab, geophysical, LAGEOS, LLR, planetary, and Galactic Center) are adapted from Refs.~\cite{Lee2017_GC,Adelberger2009,Konopliv2011,Hees2017,Borka2013,Psaltis2016,Boehle2016,Gillessen2017}.
The sensitivity is evaluated at the equivalent integration time $T_{\rm eq}$ for each configuration and is therefore limited by systematic uncertainties.
}
\label{fig:alpha_final}
\end{figure}

Here we assume the CHRONOS sensitivity at 0.5 Hz,
$S_h^{1/2} \simeq 10^{-17}\,{\rm Hz}^{-1/2}$,
as presented in~\cite{Inoue2025_CHRONOS,Inoue2025_CHRONOS_Optics,inoue2026chronosscienceprogram}.
Because the statistical sensitivity is evaluated at $T_{\rm eq}$, the
projected limit in Fig.~\ref{fig:alpha_final} already includes the
effect of the residual Newtonian systematic floor.

The convergence properties of the Yukawa expansion with respect to the
truncation order $n$ are discussed in Appendix~\ref{app:n_convergence}.
In the present analysis, we truncate the series at $n=4$.

Figure~\ref{fig:alpha_final} shows the resulting sensitivity curves
together with existing experimental and astrophysical constraints on
Yukawa-type deviations from Newtonian gravity
\cite{Adelberger2009,Fischbach1992,Konopliv2011,Hees2017,Borka2013,Psaltis2016,Boehle2016,Gillessen2017,Lee2017_GC}.
We find that the experiment exhibits strong sensitivity at large
interaction range $\lambda$.
This behavior arises because the Yukawa potential asymptotically
approaches a Newtonian-like interaction in the limit $\lambda \to \infty$.
Even in this regime, however, a measurable residual remains due to the
difference between the $L$ and $S$ source configurations, allowing the
differential setup to retain sensitivity to long-range deviations.

This feature becomes visible only when higher-order contributions are
properly included.
Previous studies, such as Raffai et al.~\cite{Raffai2011_PRD}, mainly
emphasized leading-order terms, whereas the inclusion of higher-order
geometric contributions reveals an additional sensitivity in the
large-$\lambda$ regime.
A detailed discussion of this asymptotic limit is provided in
Appendix~\ref{app:large_lambda}.
Importantly, even for very large $\lambda$, meaningful constraints can be
derived from the residual signal, which constitutes a unique aspect of
the present work.

Combining Table~\ref{tab:bscan} with Fig.~\ref{fig:alpha_final}, we find
that the optimal configuration is determined by the competition between
signal enhancement and systematic growth.
Increasing $b_L$ strengthens the Yukawa response, but it also shortens
the equivalent integration time by raising the residual Newtonian floor.
The final projected constraint is therefore obtained only after both
effects are combined through Eq.~\eqref{eq:alpha_lim}.

In particular, the best sensitivity is achieved at
\begin{equation}
|\alpha_Y| = 2.4\times10^{-5}
\quad \text{at} \quad \lambda = 8\,\mathrm{m},
\end{equation}
evaluated at the equivalent integration time
\begin{equation}
T_{\rm eq} \simeq 9.25\times10^{4}\,\mathrm{s}
\;\; (\simeq 25.7\,\mathrm{hours}).
\end{equation}

The optimal interaction scale $\lambda \sim 8\,\mathrm{m}$ originates
from the characteristic geometry of the setup, in particular the source
height $h_S$, which sets the effective distance scale of the interaction.
The input parameter uncertainties used in the error budget are adopted
from previous studies~\cite{Inoue2018_GCal,inoue2026improvingcalibrationaccuracytorque}.

We also note that a characteristic feature appears around
$\lambda \sim 2.5\,\mathrm{m}$, where the residual contributions from the
large and small GCals change sign.
This sign reversal leads to a partial cancellation of the Yukawa signal
and produces a corresponding structure in the sensitivity curve.

In this sense, the sensitivity curves presented here summarize the
maximum achievable constraint for each geometric configuration once the
available statistical integration has been exhausted up to the
systematic-error limit.

\section{Discussion}

In this work, we have proposed a method to probe Yukawa-type deviations
from Newtonian gravity using a differential gravitational calibrator
(GCal) configuration in a torsion-bar detector.
A key feature of this approach is that the leading Newtonian contribution
can be suppressed through geometric cancellation, while a residual signal
remains for non-Newtonian interactions.
In this sense, the present method transforms a calibration system into a
direct probe of new physics~\cite{Raffai2011_PRD,Inoue2018_GCal,Matone2007_CQG,inoue2026improvingcalibrationaccuracytorque}.

Compared to conventional experiments searching for short-range
modifications of gravity, such as torsion-balance measurements and atom
interferometry, our approach operates in a distinct regime
~\cite{Raffai2011_PRD,Forward1967_DynamicGravity,Hirakawa1980,Ogawa1982,Kuroda1985,Astone1991,Astone1998}.
The measurement is performed dynamically at a well-defined frequency,
rather than as a static force measurement, and directly probes the
torque acting on the rotational degree of freedom.
This frequency-domain approach provides complementary sensitivity and
offers a new avenue for exploring deviations from the inverse-square law.

The sensitivity to the Yukawa range $(\lambda)$ is determined by the
geometric configuration of the system.
In particular, the signal is maximized when $\lambda$ is comparable to
the characteristic distances of the apparatus.
In the present configuration, the optimal sensitivity is achieved around
$\lambda \sim 8\,\mathrm{m}$, which is set primarily by the source height
$h_S$ that defines the effective interaction scale.
For $\lambda \ll \mathcal{R}_S$, the Yukawa interaction is exponentially
suppressed, while for $\lambda \gg \mathcal{R}_L$, the interaction
approaches the Newtonian limit.
Nevertheless, even in the large-$\lambda$ regime, a finite residual
remains due to the asymmetry between the $L$ and $S$ configurations,
allowing the differential setup to retain sensitivity.

A central result of this work is that the achievable sensitivity is not
limited by statistical noise, but by systematic uncertainty arising from
imperfect cancellation of the Newtonian torque, consistent with general
calibration requirements in gravitational-wave detectors
~\cite{Hall2017CalibrationRequirements}.
We find that the statistical uncertainty reaches the systematic floor
within an equivalent integration time of
\begin{equation}
T_{\rm eq} \simeq 9.25\times10^{4}\,\mathrm{s}
\;\; (\simeq 25.7\,\mathrm{hours}),
\end{equation}
beyond which further integration does not improve the constraint on
$\alpha_Y$.
This establishes that the experiment operates in a
systematics-dominated regime on timescales of order one day.

The error budget analysis shows that the dominant contribution arises
from uncertainties in the geometric parameters of the source masses,
in particular $b_S$ and $b_L$, while all other sources are negligible.
The input parameter uncertainties adopted in this analysis are based on
previous calibration studies~\cite{Inoue2018_GCal,inoue2026improvingcalibrationaccuracytorque}.
This identifies precision control of the mass distribution and geometry
as the primary requirement for improving sensitivity.

We also find that the sensitivity curve exhibits a nontrivial structure
around $\lambda \sim 2.5\,\mathrm{m}$.
This feature originates from a sign change in the residual contributions
from the large and small GCals, leading to a partial cancellation of the
Yukawa signal.
Such behavior highlights the importance of accurately modeling the full
geometry of the system and cannot be captured by simplified
leading-order approximations.

An important implication of the frequency-domain operation is that the
characteristic size of the calibration system is not limited by static
constraints.
At an operating frequency of $\sim 0.5\,\mathrm{Hz}$, the required
rotational speed is sufficiently low that larger rotor radii,
$b_L \sim 0.1$--$0.5\,\mathrm{m}$, can be implemented without introducing
significant dynamical limitations, as discussed in torsion-bar detector
studies~\cite{Ando2010}.

This has two important consequences.
First, increasing the geometric scale enhances the Yukawa signal,
since the torque scales with higher powers of $b$.
Second, it shifts the optimal sensitivity toward larger values of
$\lambda$, thereby extending the accessible parameter space.
However, as demonstrated in this work, such scaling also increases the
residual Newtonian background and shortens the effective integration
time.
Therefore, the optimal design must be determined by balancing signal
enhancement against systematic growth.

Another advantage of the present formulation is that it does not rely on
a truncated multipole expansion.
The exact expression for the Yukawa-induced torque includes contributions
from all orders, ensuring the validity of the result even in regimes
where low-order approximations break down.
We have verified that the expansion converges rapidly and that truncation
at finite order does not affect the final sensitivity within the
systematic uncertainty.

Although the present work focuses on Yukawa-type interactions, the method
can be extended to a broader class of non-Newtonian forces.
Any deviation from the inverse-square law with a nontrivial distance
dependence can, in principle, be probed using a similar differential
configuration.

Future work will include a detailed evaluation of systematic
uncertainties under realistic experimental conditions
~\cite{Tanabe2025_CHRONOS_Intensity}, as well as optimization of the
calibration geometry.
In particular, improving the control of geometric parameters and
refining the differential configuration will be key to achieving
significantly stronger constraints on deviations from Newtonian gravity.

\section{Conclusion}

In this work, we have developed a quantitative framework for probing
Yukawa-type deviations from Newtonian gravity using a differential
gravitational calibrator coupled to a torsion-bar detector
~\cite{Raffai2011_PRD,Inoue2018_GCal}.

We derived the Yukawa-induced torque based on an exact formulation of
the gravitational interaction and expressed it as a function of the
interaction parameters $(\alpha_Y,\lambda)$.
The resulting torque was converted into a strain-equivalent signal
through the mechanical response of the torsion-bar system
~\cite{Saulson1990,Ando2010}, enabling a direct comparison with the
detector noise.

Using this framework, we evaluated the signal-to-noise ratio and
constructed projected sensitivity curves in the
$(\alpha_Y,\lambda)$ parameter space.
Our main result is that the experiment reaches a projected sensitivity of
\begin{equation}
|\alpha_Y| = 2.4\times10^{-5}
\quad \text{at} \quad \lambda = 8\,\mathrm{m},
\end{equation}
evaluated at the equivalent integration time
\begin{equation}
T_{\rm eq} \simeq 9.25\times10^{4}\,\mathrm{s}.
\end{equation}
\begin{equation}
T_{\rm eq} \simeq 9.25\times10^{4}\,\mathrm{s}
\;\; (\simeq 25.7\,\mathrm{hours}).
\end{equation}
This corresponds to approximately one day ($\sim 26$ hours),
indicating that further integration does not improve the sensitivity
and that the experiment operates in a systematics-dominated regime.

A central finding of this study is that the sensitivity is not limited
by statistical noise, but by systematic uncertainty arising from
imperfect cancellation of the Newtonian torque, consistent with
calibration limits in gravitational-wave detectors
~\cite{Hall2017CalibrationRequirements,Cahillane2017_Calibration}.
The statistical noise reaches the systematic floor within
$\sim 10$ hours, beyond which further integration does not improve the
constraint.
This establishes that the experiment operates in a
systematics-dominated regime.

The error budget analysis shows that the residual uncertainty is
overwhelmingly dominated by the geometric parameters of the source
masses, in particular $b_S$ and $b_L$, each contributing nearly $50\%$
of the total variance.
In contrast, uncertainties in the gravitational constant and detector
parameters are negligible.
This identifies precision control of the source geometry as the key
requirement for improving experimental sensitivity.

From the theoretical side, we have demonstrated that the Yukawa torque
can be expressed as a unified expansion that consistently reduces to
the Newtonian limit.
The convergence of the series is rapid due to the small geometric
parameter $(b/R)^2$, and truncation at finite order introduces an error
well below the systematic uncertainty.
Thus, the theoretical prediction can be regarded as effectively exact
within the experimental precision.

A particularly important and unique aspect of this work is that the
differential configuration retains sensitivity even in the large-$\lambda$
limit.
While the Yukawa interaction asymptotically approaches Newtonian gravity
for $\lambda \to \infty$, a measurable residual signal remains due to
the geometric asymmetry between the $L$ and $S$ configurations.
This allows meaningful constraints to be placed even at very large
interaction ranges, a feature that has not been fully exploited in
previous studies.

The resulting sensitivity curve exhibits three distinct regimes:
exponential suppression at short interaction ranges,
optimal sensitivity at $\lambda \sim \mathcal{O}(1\,\mathrm{m})$, and
a finite asymptotic sensitivity at large $\lambda$.
This structure reflects the interplay between Yukawa suppression,
geometric enhancement, and incomplete Newtonian cancellation.

In summary, the framework presented here establishes a direct and
quantitative connection between detector performance, calibration
systematics, and constraints on non-Newtonian gravity.
The key result,
$|\alpha_Y| = 2.4\times10^{-5}$ at $\lambda=8\,\mathrm{m}$,
demonstrates that sub-Hz torsion-bar detectors provide a powerful and
complementary probe of gravity at meter-scale interaction ranges.

Future improvements will focus on reducing geometric uncertainties and
optimizing the differential configuration, which will directly translate
into stronger constraints on Yukawa-type deviations from Newtonian
gravity.


\acknowledgments
We would like to express our sincere gratitude to Y-C.Lin, M.Hasegawa, T.Kanayama and M.Hazumi for their valuable discussions and continuous support throughout this work. We also acknowledge the support and collaborative environment provided by the Department of Physics and the Center for High Energy and High Field (CHiP) at National Central University, the Institute of Physics, Academia Sinica, the National Institute of Physics, University of the Philippines Diliman. Y.I. is supported by the National Science and Technology Council (NSTC) of Taiwan under Grant No. 114-2112-M-008-006, and by Academia Sinica under Grant No. AS-TP-112-M01.
\appendix
\section{Newtonian limit}
\label{app:newton}

Before proceeding, it is useful to verify that the Yukawa expansion
correctly reduces to the Newtonian expression in the limit $\beta \to 0$.

From the definition of the Bessel polynomials~\cite{Grosswald1978}
(see Eq.~\eqref{eq:ck_bessel_poly}), we have
\begin{equation}
y_k(0) = \frac{(2k)!}{2^k k!}.
\end{equation}
Therefore,
\begin{equation}
\frac{y_k(0)}{2^k k!}
=
\frac{(2k)!}{4^k (k!)^2}
=
\frac{1}{4^k}\binom{2k}{k}.
\label{eq:newton_limit_coeff}
\end{equation}

Substituting this into the Yukawa kernel expansion
(Eq.~\eqref{eq:kernel_definition}), we obtain
\begin{equation}
\frac{1}{\sqrt{1-u}}
=
\sum_{k=0}^{\infty}
\frac{1}{4^k}\binom{2k}{k} u^k,
\end{equation}
which is exactly the standard Newtonian expansion.

We now consider the coefficients appearing in the torque expansion.
The Newtonian coefficient can be written as
\begin{equation}
A_n
=
\frac{1}{2^{4n-1}}
\binom{4n}{2n}
\binom{2n}{n-1}.
\label{eq:An_def}
\end{equation}

Using Eq.~\eqref{eq:newton_limit_coeff}, the Yukawa coefficient
$B_n(\beta)$ can be expressed in a compact form as
\begin{equation}
B_n(\beta)
=
A_n \frac{y_{2n}(\beta)}{y_{2n}(0)}.
\label{eq:Bn_from_An}
\end{equation}

This relation makes the structure of the expansion transparent:
the Yukawa interaction preserves the Newtonian geometrical
dependence encoded in $A_n$, while modifying only the radial
dependence through the ratio $y_{2n}(\beta)/y_{2n}(0)$.

In the limit $\beta \to 0$, we have $y_{2n}(\beta) \to y_{2n}(0)$,
and therefore
\begin{equation}
B_n(\beta) \to A_n,
\label{eq:lim_Bn_An}
\end{equation}
showing that the Yukawa expansion smoothly reduces to the
Newtonian one.
\section{Mechanical transfer function of the torsion-bar mode}
\label{app:transfer_function}

In this appendix, we derive the mechanical transfer function used to
convert an external gravitational torque into the corresponding angular
response of the torsion bar.

We consider the rotational degree of freedom $\theta(t)$ of a torsion
bar with moment of inertia $I$, torsional spring constant $\kappa$,
and viscous damping coefficient $\Gamma$.
The equation of motion under an external torque $\tau(t)$ is
\begin{equation}
I\ddot{\theta}(t)+\Gamma\dot{\theta}(t)+\kappa\theta(t)=\tau(t),
\label{eq:eom_theta_time}
\end{equation}
which is the standard form of a damped harmonic oscillator~\cite{Saulson1990}.

Assuming harmonic time dependence,
\begin{equation}
\theta(t)=\theta(\Omega)e^{-i\Omega t},
\qquad
\tau(t)=\tau(\Omega)e^{-i\Omega t},
\label{eq:harmonic_ansatz}
\end{equation}
Eq.~(\ref{eq:eom_theta_time}) becomes
\begin{equation}
\left(
-\Omega^2 I - i\Omega\Gamma + \kappa
\right)\theta(\Omega)=\tau(\Omega).
\label{eq:eom_theta_freq}
\end{equation}
Therefore, the mechanical susceptibility is given by
\begin{equation}
\chi(\Omega)
\equiv
\frac{\theta(\Omega)}{\tau(\Omega)}
=
\frac{1}{I\left(\omega_0^2-\Omega^2\right)-i\Gamma\Omega}.
\label{eq:chi_omega0}.
\end{equation}

where the torsional resonance frequency is obtained as
\begin{equation}
\omega_0^2=\frac{\kappa}{I},
\label{eq:omega0_def}
\end{equation}

In the high-frequency regime relevant for the present study,
$(\Omega \gg \omega_0)$, the inertial term dominates over the restoring
term, and the susceptibility reduces to
\begin{equation}
\chi(\Omega)\simeq -\frac{1}{I\Omega^2},
\label{eq:chi_highf_appendix}
\end{equation}
where damping has also been neglected since
$(|\Gamma\Omega| \ll I\Omega^2)$ in this limit.

This is the transfer function used in the main text to convert the
external GCal or Yukawa torque into an angular response of the torsion
bar, consistent with standard treatments of mechanical response in
gravitational-wave detectors~\cite{Saulson1990}. The corresponding strain-equivalent signal is then obtained through
the geometrical conversion factor $(\eta_g F_{\rm eff}/2)$ introduced in Inoue et al.~\cite{Inoue2025_CHRONOS}.

\section{Convergence of the Yukawa expansion}
\label{app:n_convergence}
In this appendix, we examine the convergence of the Yukawa torque expansion
by directly comparing numerical results obtained with different truncation orders.

The Yukawa torque is evaluated as a finite sum truncated at order $n_{\rm max}$.
To assess convergence, we compute the projected sensitivity
$\alpha_{Y{\rm lim}}(\lambda)$ using truncation orders $n=1$--$4$,
while keeping all other parameters fixed to the values used in the main analysis.

\begin{figure}[t]
\centering
\includegraphics[width=0.9\linewidth]{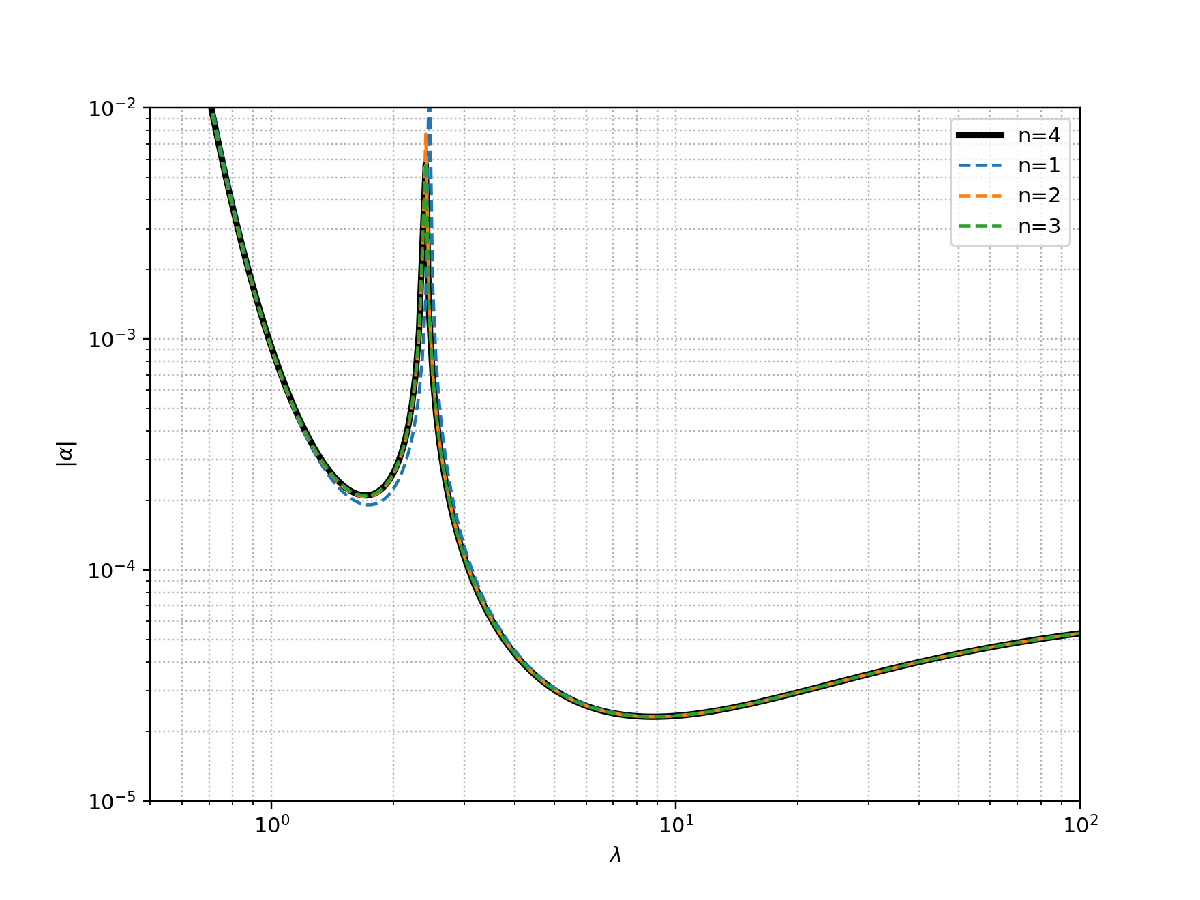}
\caption{
Sensitivity curves $\alpha_{Y{\rm lim}}(\lambda)$ evaluated with truncation orders $n=1$--$4$.
The curves progressively converge as $n$ increases.
The results for $n \geq 4$ are visually indistinguishable over the entire range of $\lambda$.
}
\label{fig:yukawa_conv}
\end{figure}

The results are shown in Fig.~\ref{fig:yukawa_conv}.

The figure shows that the sensitivity curves rapidly approach a stable solution
as the truncation order increases.
At low truncation order ($n=1$--$2$), visible deviations are present,
indicating that higher-order terms contribute non-negligibly to the signal.
However, these differences decrease systematically as $n$ increases.

For $n \geq 4$, the curves overlap within the resolution of the plot
over the full range of $\lambda$ considered.
This demonstrates that the series has effectively converged at this order.

A localized deviation between truncation orders is observed
around $\lambda \sim 0.3$--$1\,\mathrm{m}$,
where higher-order contributions temporarily modify the signal.
Nevertheless, this difference becomes negligible once $n \geq 4$,
and does not affect the final sensitivity curve.

These results demonstrate that the Yukawa expansion converges rapidly
in the present configuration.
We therefore adopt $n_{\rm max}=4$ in all numerical evaluations.

Importantly, the residual differences between successive truncation orders
are much smaller than the systematic uncertainty discussed in the main text.
Therefore, the truncation error does not limit the sensitivity of the experiment.

\section{Large-\texorpdfstring{$\lambda$}{lambda} limit of the Yukawa torque}
\label{app:large_lambda}

In this appendix, we derive the asymptotic form of the Yukawa torque in the large-$\lambda$ limit and extract the leading residual contribution.

According to Eq.(\ref{eq:tau_yukawa_lambda}), we define
\begin{equation}
F_n(\beta^{(Y)}) \equiv e^{-\beta^{(Y)}} B_n(\beta^{(Y)}).
\end{equation}

Then Eq.~(\ref{eq:tau_yukawa_lambda}) becomes
\begin{align}
\tau_Y(\lambda)
&=
2GM\alpha_Y\gamma \frac{m}{R_S}
\sum_{n=1}^{\infty}
\left(\frac{2ab}{R_S^2}\right)^{2n}
\nonumber\\
&\quad\times
\left[
2^{2n}\left(\frac{R_S}{R_L}\right)^{2n+1} F_n(\beta^{(Y)}_L)
-
F_n(\beta^{(Y)}_S)
\right]
\sin(2\omega t).
\label{eq:tauY_Fn_app}
\end{align}

In the limit $\lambda \to \infty$, we have $\beta^{(Y)}_S,\beta^{(Y)}_L \to 0$, and
\begin{equation}
F_n(\beta^{(Y)})=A_n+\mathcal{O}((\beta^{(Y)})^2).
\end{equation}

Substituting this, we obtain
\begin{align}
\tau_Y(\lambda)
&=
2GM\alpha_Y\gamma \frac{m}{R_S}
\sum_{n=1}^{\infty}
\left(\frac{2ab}{R_S^2}\right)^{2n}
\nonumber\\
&\quad\times
A_n
\left[
2^{2n}\left(\frac{R_S}{R_L}\right)^{2n+1}
-1
\right]
\sin(2\omega t)
+ \mathcal{O}(\lambda^{-2}).
\label{eq:tauY_large_lambda_app}
\end{align}

Focusing on the leading contribution ($n=1$), we obtain
\begin{align}
\tau_Y^{(n=1,\infty)}
&=
2GM\alpha_Y\gamma \frac{m}{R_S}
\left(\frac{2ab}{R_S^2}\right)^2
A_1
\nonumber\\
&\quad\times
\left[
4\left(\frac{R_S}{R_L}\right)^3-1
\right]
\sin(2\omega t).
\label{eq:n1_general_app}
\end{align}

The coefficient $A_n$ is given by
\begin{equation}
A_n
=
\frac{1}{2^{4n-1}}
\binom{4n}{2n}
\binom{2n}{n-1},
\qquad
A_1=\frac{3}{4}.
\end{equation}

Therefore,
\begin{align}
\tau_Y^{(n=1,\infty)}
&=
6GM\alpha_Y\gamma m
\frac{a^2 b^2}{R_S^5}
\nonumber\\
&\quad\times
\left[
4\left(\frac{R_S}{R_L}\right)^3-1
\right]
\sin(2\omega t).
\label{eq:n1_final_app}
\end{align}

This result shows that, unless the first Newtonian-like term is exactly canceled, a finite residual remains even in the large-$\lambda$ limit.

\bibliographystyle{apsrev4-1}
\bibliography{chronos_GCal}

\begin{thebibliography}{42}%
\makeatletter
\providecommand \@ifxundefined [1]{%
 \@ifx{#1\undefined}
}%
\providecommand \@ifnum [1]{%
 \ifnum #1\expandafter \@firstoftwo
 \else \expandafter \@secondoftwo
 \fi
}%
\providecommand \@ifx [1]{%
 \ifx #1\expandafter \@firstoftwo
 \else \expandafter \@secondoftwo
 \fi
}%
\providecommand \natexlab [1]{#1}%
\providecommand \enquote  [1]{``#1''}%
\providecommand \bibnamefont  [1]{#1}%
\providecommand \bibfnamefont [1]{#1}%
\providecommand \citenamefont [1]{#1}%
\providecommand \href@noop [0]{\@secondoftwo}%
\providecommand \href [0]{\begingroup \@sanitize@url \@href}%
\providecommand \@href[1]{\@@startlink{#1}\@@href}%
\providecommand \@@href[1]{\endgroup#1\@@endlink}%
\providecommand \@sanitize@url [0]{\catcode `\\12\catcode `\$12\catcode
  `\&12\catcode `\#12\catcode `\^12\catcode `\_12\catcode `\%12\relax}%
\providecommand \@@startlink[1]{}%
\providecommand \@@endlink[0]{}%
\providecommand \url  [0]{\begingroup\@sanitize@url \@url }%
\providecommand \@url [1]{\endgroup\@href {#1}{\urlprefix }}%
\providecommand \urlprefix  [0]{URL }%
\providecommand \Eprint [0]{\href }%
\providecommand \doibase [0]{http://dx.doi.org/}%
\providecommand \selectlanguage [0]{\@gobble}%
\providecommand \bibinfo  [0]{\@secondoftwo}%
\providecommand \bibfield  [0]{\@secondoftwo}%
\providecommand \translation [1]{[#1]}%
\providecommand \BibitemOpen [0]{}%
\providecommand \bibitemStop [0]{}%
\providecommand \bibitemNoStop [0]{.\EOS\space}%
\providecommand \EOS [0]{\spacefactor3000\relax}%
\providecommand \BibitemShut  [1]{\csname bibitem#1\endcsname}%
\let\auto@bib@innerbib\@empty
\bibitem [{\citenamefont {Fischbach}\ and\ \citenamefont
  {Talmadge}(1999)}]{Fischbach1999}%
  \BibitemOpen
  \bibfield  {author} {\bibinfo {author} {\bibfnamefont {E.}~\bibnamefont
  {Fischbach}}\ and\ \bibinfo {author} {\bibfnamefont {C.}~\bibnamefont
  {Talmadge}},\ }\href {\doibase 10.1007/978-1-4612-0527-2} {\emph {\bibinfo
  {title} {The Search for Non-Newtonian Gravity}}}\ (\bibinfo  {publisher}
  {Springer},\ \bibinfo {year} {1999})\BibitemShut {NoStop}%
\bibitem [{\citenamefont {Adelberger}\ \emph {et~al.}(2003)\citenamefont
  {Adelberger} \emph {et~al.}}]{Adelberger2003}%
  \BibitemOpen
  \bibfield  {author} {\bibinfo {author} {\bibfnamefont {E.~G.}\ \bibnamefont
  {Adelberger}} \emph {et~al.},\ }\href {\doibase
  10.1146/annurev.nucl.53.041002.110503} {\bibfield  {journal} {\bibinfo
  {journal} {Ann. Rev. Nucl. Part. Sci.}\ }\textbf {\bibinfo {volume} {53}},\
  \bibinfo {pages} {77} (\bibinfo {year} {2003})}\BibitemShut {NoStop}%
\bibitem [{\citenamefont {Adelberger}\ \emph {et~al.}(2009)\citenamefont
  {Adelberger} \emph {et~al.}}]{Adelberger2009}%
  \BibitemOpen
  \bibfield  {author} {\bibinfo {author} {\bibfnamefont {E.~G.}\ \bibnamefont
  {Adelberger}} \emph {et~al.},\ }\href {\doibase 10.1016/j.ppnp.2008.08.002}
  {\bibfield  {journal} {\bibinfo  {journal} {Prog. Part. Nucl. Phys.}\
  }\textbf {\bibinfo {volume} {62}},\ \bibinfo {pages} {102} (\bibinfo {year}
  {2009})}\BibitemShut {NoStop}%
\bibitem [{\citenamefont {Kapner}\ \emph {et~al.}(2007)\citenamefont {Kapner}
  \emph {et~al.}}]{Kapner2007}%
  \BibitemOpen
  \bibfield  {author} {\bibinfo {author} {\bibfnamefont {D.~J.}\ \bibnamefont
  {Kapner}} \emph {et~al.},\ }\href {\doibase 10.1103/PhysRevLett.98.021101}
  {\bibfield  {journal} {\bibinfo  {journal} {Phys. Rev. Lett.}\ }\textbf
  {\bibinfo {volume} {98}},\ \bibinfo {pages} {021101} (\bibinfo {year}
  {2007})},\ \Eprint {http://arxiv.org/abs/hep-ph/0611184}
  {arXiv:hep-ph/0611184} \BibitemShut {NoStop}%
\bibitem [{\citenamefont {Geraci}\ \emph {et~al.}(2008)\citenamefont {Geraci}
  \emph {et~al.}}]{Geraci2008}%
  \BibitemOpen
  \bibfield  {author} {\bibinfo {author} {\bibfnamefont {A.~A.}\ \bibnamefont
  {Geraci}} \emph {et~al.},\ }\href {\doibase 10.1103/PhysRevD.78.022002}
  {\bibfield  {journal} {\bibinfo  {journal} {Phys. Rev. D}\ }\textbf {\bibinfo
  {volume} {78}},\ \bibinfo {pages} {022002} (\bibinfo {year}
  {2008})}\BibitemShut {NoStop}%
\bibitem [{\citenamefont {Raffai}\ \emph {et~al.}(2011)\citenamefont {Raffai}
  \emph {et~al.}}]{Raffai2011_PRD}%
  \BibitemOpen
  \bibfield  {author} {\bibinfo {author} {\bibfnamefont {P.}~\bibnamefont
  {Raffai}} \emph {et~al.},\ }\href {\doibase 10.1103/PhysRevD.84.082002}
  {\bibfield  {journal} {\bibinfo  {journal} {Phys. Rev. D}\ }\textbf {\bibinfo
  {volume} {84}},\ \bibinfo {pages} {082002} (\bibinfo {year}
  {2011})}\BibitemShut {NoStop}%
\bibitem [{\citenamefont {Konopliv}\ \emph {et~al.}(2011)\citenamefont
  {Konopliv} \emph {et~al.}}]{Konopliv2011}%
  \BibitemOpen
  \bibfield  {author} {\bibinfo {author} {\bibfnamefont {A.~S.}\ \bibnamefont
  {Konopliv}} \emph {et~al.},\ }\href {\doibase 10.1016/j.icarus.2010.10.004}
  {\bibfield  {journal} {\bibinfo  {journal} {Icarus}\ }\textbf {\bibinfo
  {volume} {211}},\ \bibinfo {pages} {401} (\bibinfo {year}
  {2011})}\BibitemShut {NoStop}%
\bibitem [{\citenamefont {Hees}\ \emph {et~al.}(2017)\citenamefont {Hees} \emph
  {et~al.}}]{Hees2017}%
  \BibitemOpen
  \bibfield  {author} {\bibinfo {author} {\bibfnamefont {A.}~\bibnamefont
  {Hees}} \emph {et~al.},\ }\href {\doibase 10.1103/PhysRevLett.118.211101}
  {\bibfield  {journal} {\bibinfo  {journal} {Phys. Rev. Lett.}\ }\textbf
  {\bibinfo {volume} {118}},\ \bibinfo {pages} {211101} (\bibinfo {year}
  {2017})},\ \Eprint {http://arxiv.org/abs/1705.07902} {arXiv:1705.07902}
  \BibitemShut {NoStop}%
\bibitem [{\citenamefont {Lee}\ \emph {et~al.}(2017)\citenamefont {Lee} \emph
  {et~al.}}]{Lee2017_GC}%
  \BibitemOpen
  \bibfield  {author} {\bibinfo {author} {\bibfnamefont {J.~G.}\ \bibnamefont
  {Lee}} \emph {et~al.},\ }\href {\doibase 10.1103/PhysRevLett.118.211101}
  {\bibfield  {journal} {\bibinfo  {journal} {Phys. Rev. Lett.}\ }\textbf
  {\bibinfo {volume} {118}},\ \bibinfo {pages} {211101} (\bibinfo {year}
  {2017})},\ \Eprint {http://arxiv.org/abs/1705.10792} {arXiv:1705.10792
  [gr-qc]} \BibitemShut {NoStop}%
\bibitem [{\citenamefont {Abbott}\ \emph {et~al.}(2016)\citenamefont {Abbott}
  \emph {et~al.}}]{LIGO2016}%
  \BibitemOpen
  \bibfield  {author} {\bibinfo {author} {\bibfnamefont {B.~P.}\ \bibnamefont
  {Abbott}} \emph {et~al.},\ }\href {\doibase 10.1103/PhysRevLett.116.061102}
  {\bibfield  {journal} {\bibinfo  {journal} {Phys. Rev. Lett.}\ }\textbf
  {\bibinfo {volume} {116}},\ \bibinfo {pages} {061102} (\bibinfo {year}
  {2016})}\BibitemShut {NoStop}%
\bibitem [{\citenamefont {Abbott}\ \emph {et~al.}(2019)\citenamefont {Abbott}
  \emph {et~al.}}]{GWTC1}%
  \BibitemOpen
  \bibfield  {author} {\bibinfo {author} {\bibfnamefont {B.~P.}\ \bibnamefont
  {Abbott}} \emph {et~al.},\ }\href {\doibase 10.1103/PhysRevX.9.031040}
  {\bibfield  {journal} {\bibinfo  {journal} {Phys. Rev. X}\ }\textbf {\bibinfo
  {volume} {9}},\ \bibinfo {pages} {031040} (\bibinfo {year}
  {2019})}\BibitemShut {NoStop}%
\bibitem [{\citenamefont {Abbott}\ \emph {et~al.}(2021)\citenamefont {Abbott}
  \emph {et~al.}}]{GWTC2}%
  \BibitemOpen
  \bibfield  {author} {\bibinfo {author} {\bibfnamefont {R.}~\bibnamefont
  {Abbott}} \emph {et~al.},\ }\href {\doibase 10.1103/PhysRevX.11.021053}
  {\bibfield  {journal} {\bibinfo  {journal} {Phys. Rev. X}\ }\textbf {\bibinfo
  {volume} {11}},\ \bibinfo {pages} {021053} (\bibinfo {year}
  {2021})}\BibitemShut {NoStop}%
\bibitem [{\citenamefont {Abbott}\ \emph {et~al.}(2023)\citenamefont {Abbott}
  \emph {et~al.}}]{GWTC3}%
  \BibitemOpen
  \bibfield  {author} {\bibinfo {author} {\bibfnamefont {R.}~\bibnamefont
  {Abbott}} \emph {et~al.},\ }\href {\doibase 10.1103/PhysRevX.13.041039}
  {\bibfield  {journal} {\bibinfo  {journal} {Phys. Rev. X}\ }\textbf {\bibinfo
  {volume} {13}},\ \bibinfo {pages} {041039} (\bibinfo {year}
  {2023})}\BibitemShut {NoStop}%
\bibitem [{\citenamefont {Ando}\ \emph {et~al.}(2010)\citenamefont {Ando} \emph
  {et~al.}}]{Ando2010}%
  \BibitemOpen
  \bibfield  {author} {\bibinfo {author} {\bibfnamefont {M.}~\bibnamefont
  {Ando}} \emph {et~al.},\ }\href {\doibase 10.1103/PhysRevLett.105.161101}
  {\bibfield  {journal} {\bibinfo  {journal} {Phys. Rev. Lett.}\ }\textbf
  {\bibinfo {volume} {105}},\ \bibinfo {pages} {161101} (\bibinfo {year}
  {2010})}\BibitemShut {NoStop}%
\bibitem [{\citenamefont {Ju}\ \emph {et~al.}(2019)\citenamefont {Ju},
  \citenamefont {Zhao},\ and\ \citenamefont {Blair}}]{TorPeDO2019}%
  \BibitemOpen
  \bibfield  {author} {\bibinfo {author} {\bibfnamefont {L.}~\bibnamefont
  {Ju}}, \bibinfo {author} {\bibfnamefont {C.}~\bibnamefont {Zhao}}, \ and\
  \bibinfo {author} {\bibfnamefont {D.}~\bibnamefont {Blair}},\ }in\ \href
  {\doibase 10.1142/9789813226609_0292} {\emph {\bibinfo {booktitle} {Proc.
  13th Marcel Grossmann Meeting}}}\ (\bibinfo {year} {2019})\ pp.\ \bibinfo
  {pages} {2487--2492}\BibitemShut {NoStop}%
\bibitem [{\citenamefont {Inoue}\ \emph
  {et~al.}(2026{\natexlab{a}})\citenamefont {Inoue} \emph
  {et~al.}}]{Inoue2025_CHRONOS}%
  \BibitemOpen
  \bibfield  {author} {\bibinfo {author} {\bibfnamefont {Y.}~\bibnamefont
  {Inoue}} \emph {et~al.},\ }\href@noop {} {\  (\bibinfo {year}
  {2026}{\natexlab{a}})},\ \Eprint {http://arxiv.org/abs/2509.23172}
  {arXiv:2509.23172 [astro-ph.IM]} \BibitemShut {NoStop}%
\bibitem [{\citenamefont {Inoue}\ \emph {et~al.}(2025)\citenamefont {Inoue},
  \citenamefont {Tanabe}, \citenamefont {Ismail}, \citenamefont {Kumar},
  \citenamefont {Onglao},\ and\ \citenamefont {Yu}}]{Inoue2025_CHRONOS_Optics}%
  \BibitemOpen
  \bibfield  {author} {\bibinfo {author} {\bibfnamefont {Y.}~\bibnamefont
  {Inoue}}, \bibinfo {author} {\bibfnamefont {D.}~\bibnamefont {Tanabe}},
  \bibinfo {author} {\bibfnamefont {M.~A.}\ \bibnamefont {Ismail}}, \bibinfo
  {author} {\bibfnamefont {V.}~\bibnamefont {Kumar}}, \bibinfo {author}
  {\bibfnamefont {M.~J.~S.}\ \bibnamefont {Onglao}}, \ and\ \bibinfo {author}
  {\bibfnamefont {T.-C.}\ \bibnamefont {Yu}},\ }\href {\doibase
  10.48550/arXiv.2510.24780} {\  (\bibinfo {year} {2025}),\
  10.48550/arXiv.2510.24780},\ \Eprint {http://arxiv.org/abs/2510.24780}
  {arXiv:2510.24780 [physics.ins-det]} \BibitemShut {NoStop}%
\bibitem [{\citenamefont {Inoue}\ \emph
  {et~al.}(2026{\natexlab{b}})\citenamefont {Inoue} \emph
  {et~al.}}]{inoue2026chronosscienceprogram}%
  \BibitemOpen
  \bibfield  {author} {\bibinfo {author} {\bibfnamefont {Y.}~\bibnamefont
  {Inoue}} \emph {et~al.},\ }\href@noop {} {\  (\bibinfo {year}
  {2026}{\natexlab{b}})},\ \Eprint {http://arxiv.org/abs/2603.10070}
  {arXiv:2603.10070 [astro-ph.IM]} \BibitemShut {NoStop}%
\bibitem [{\citenamefont {Saulson}(1994)}]{Saulson1990}%
  \BibitemOpen
  \bibfield  {author} {\bibinfo {author} {\bibfnamefont {P.~R.}\ \bibnamefont
  {Saulson}},\ }\href {\doibase 10.1142/2927} {\emph {\bibinfo {title}
  {Fundamentals of Interferometric Gravitational Wave Detectors}}}\ (\bibinfo
  {publisher} {World Scientific},\ \bibinfo {year} {1994})\BibitemShut
  {NoStop}%
\bibitem [{\citenamefont {Karki}\ \emph {et~al.}(2016)\citenamefont {Karki}
  \emph {et~al.}}]{Karki2016}%
  \BibitemOpen
  \bibfield  {author} {\bibinfo {author} {\bibfnamefont {S.}~\bibnamefont
  {Karki}} \emph {et~al.},\ }\href {\doibase 10.1063/1.4967303} {\bibfield
  {journal} {\bibinfo  {journal} {Rev. Sci. Instrum.}\ }\textbf {\bibinfo
  {volume} {87}},\ \bibinfo {pages} {114503} (\bibinfo {year}
  {2016})}\BibitemShut {NoStop}%
\bibitem [{\citenamefont {Goetz}\ \emph {et~al.}(2010)\citenamefont {Goetz}
  \emph {et~al.}}]{Goetz2010_LIGO_Calibration}%
  \BibitemOpen
  \bibfield  {author} {\bibinfo {author} {\bibfnamefont {E.}~\bibnamefont
  {Goetz}} \emph {et~al.},\ }\href {\doibase 10.1088/0264-9381/27/8/084024}
  {\bibfield  {journal} {\bibinfo  {journal} {Class. Quantum Grav.}\ }\textbf
  {\bibinfo {volume} {27}},\ \bibinfo {pages} {084024} (\bibinfo {year}
  {2010})}\BibitemShut {NoStop}%
\bibitem [{\citenamefont {Inoue}\ \emph {et~al.}(2023)\citenamefont {Inoue}
  \emph {et~al.}}]{Inoue2023_KAGRA_PCal}%
  \BibitemOpen
  \bibfield  {author} {\bibinfo {author} {\bibfnamefont {Y.}~\bibnamefont
  {Inoue}} \emph {et~al.},\ }\href {\doibase 10.1063/5.0139121} {\bibfield
  {journal} {\bibinfo  {journal} {Rev. Sci. Instrum.}\ } (\bibinfo {year}
  {2023}),\ 10.1063/5.0139121},\ \Eprint {http://arxiv.org/abs/2302.12180}
  {arXiv:2302.12180 [gr-qc]} \BibitemShut {NoStop}%
\bibitem [{\citenamefont {Est{\'e}vez}\ \emph {et~al.}(2018)\citenamefont
  {Est{\'e}vez} \emph {et~al.}}]{Estevez2018NCal}%
  \BibitemOpen
  \bibfield  {author} {\bibinfo {author} {\bibfnamefont {D.}~\bibnamefont
  {Est{\'e}vez}} \emph {et~al.},\ }\href {\doibase 10.1088/1361-6382/aae6c1}
  {\bibfield  {journal} {\bibinfo  {journal} {Class. Quantum Grav.}\ }\textbf
  {\bibinfo {volume} {35}},\ \bibinfo {pages} {235009} (\bibinfo {year}
  {2018})}\BibitemShut {NoStop}%
\bibitem [{\citenamefont {Acernese}\ \emph {et~al.}(2018)\citenamefont
  {Acernese} \emph {et~al.}}]{Acernese2018VirgoNCal}%
  \BibitemOpen
  \bibfield  {author} {\bibinfo {author} {\bibfnamefont {F.}~\bibnamefont
  {Acernese}} \emph {et~al.},\ }\href {\doibase 10.1088/1361-6382/aadf0e}
  {\bibfield  {journal} {\bibinfo  {journal} {Class. Quantum Grav.}\ }\textbf
  {\bibinfo {volume} {35}},\ \bibinfo {pages} {205004} (\bibinfo {year}
  {2018})}\BibitemShut {NoStop}%
\bibitem [{\citenamefont {Forward}\ and\ \citenamefont
  {Miller}(1967)}]{Forward1967_DynamicGravity}%
  \BibitemOpen
  \bibfield  {author} {\bibinfo {author} {\bibfnamefont {R.~L.}\ \bibnamefont
  {Forward}}\ and\ \bibinfo {author} {\bibfnamefont {L.~R.}\ \bibnamefont
  {Miller}},\ }\href {\doibase 10.1063/1.1709055} {\bibfield  {journal}
  {\bibinfo  {journal} {J. Appl. Phys.}\ }\textbf {\bibinfo {volume} {38}},\
  \bibinfo {pages} {512} (\bibinfo {year} {1967})}\BibitemShut {NoStop}%
\bibitem [{\citenamefont {Hirakawa}\ \emph {et~al.}(1980)\citenamefont
  {Hirakawa}, \citenamefont {Tsubono},\ and\ \citenamefont
  {Oide}}]{Hirakawa1980}%
  \BibitemOpen
  \bibfield  {author} {\bibinfo {author} {\bibfnamefont {H.}~\bibnamefont
  {Hirakawa}}, \bibinfo {author} {\bibfnamefont {K.}~\bibnamefont {Tsubono}}, \
  and\ \bibinfo {author} {\bibfnamefont {K.}~\bibnamefont {Oide}},\ }\href
  {\doibase 10.1038/283184a0} {\bibfield  {journal} {\bibinfo  {journal}
  {Nature}\ }\textbf {\bibinfo {volume} {283}},\ \bibinfo {pages} {184}
  (\bibinfo {year} {1980})}\BibitemShut {NoStop}%
\bibitem [{\citenamefont {Ogawa}\ \emph {et~al.}(1982)\citenamefont {Ogawa},
  \citenamefont {Tsubono},\ and\ \citenamefont {Hirakawa}}]{Ogawa1982}%
  \BibitemOpen
  \bibfield  {author} {\bibinfo {author} {\bibfnamefont {Y.}~\bibnamefont
  {Ogawa}}, \bibinfo {author} {\bibfnamefont {K.}~\bibnamefont {Tsubono}}, \
  and\ \bibinfo {author} {\bibfnamefont {H.}~\bibnamefont {Hirakawa}},\ }\href
  {\doibase 10.1103/PhysRevD.26.729} {\bibfield  {journal} {\bibinfo  {journal}
  {Phys. Rev. D}\ }\textbf {\bibinfo {volume} {26}},\ \bibinfo {pages} {729}
  (\bibinfo {year} {1982})}\BibitemShut {NoStop}%
\bibitem [{\citenamefont {Kuroda}\ and\ \citenamefont
  {Hirakawa}(1985)}]{Kuroda1985}%
  \BibitemOpen
  \bibfield  {author} {\bibinfo {author} {\bibfnamefont {K.}~\bibnamefont
  {Kuroda}}\ and\ \bibinfo {author} {\bibfnamefont {H.}~\bibnamefont
  {Hirakawa}},\ }\href {\doibase 10.1103/PhysRevD.32.342} {\bibfield  {journal}
  {\bibinfo  {journal} {Phys. Rev. D}\ }\textbf {\bibinfo {volume} {32}},\
  \bibinfo {pages} {342} (\bibinfo {year} {1985})}\BibitemShut {NoStop}%
\bibitem [{\citenamefont {Astone}\ \emph {et~al.}(1991)\citenamefont {Astone}
  \emph {et~al.}}]{Astone1991}%
  \BibitemOpen
  \bibfield  {author} {\bibinfo {author} {\bibfnamefont {P.}~\bibnamefont
  {Astone}} \emph {et~al.},\ }\href {\doibase 10.1007/BF01548556} {\bibfield
  {journal} {\bibinfo  {journal} {Z. Phys. C}\ }\textbf {\bibinfo {volume}
  {50}},\ \bibinfo {pages} {21} (\bibinfo {year} {1991})}\BibitemShut {NoStop}%
\bibitem [{\citenamefont {Astone}\ \emph {et~al.}(1998)\citenamefont {Astone}
  \emph {et~al.}}]{Astone1998}%
  \BibitemOpen
  \bibfield  {author} {\bibinfo {author} {\bibfnamefont {P.}~\bibnamefont
  {Astone}} \emph {et~al.},\ }\href {\doibase 10.1007/s100520050307} {\bibfield
   {journal} {\bibinfo  {journal} {Eur. Phys. J. C}\ }\textbf {\bibinfo
  {volume} {5}},\ \bibinfo {pages} {651} (\bibinfo {year} {1998})}\BibitemShut
  {NoStop}%
\bibitem [{\citenamefont {Inoue}\ \emph {et~al.}(2018)\citenamefont {Inoue}
  \emph {et~al.}}]{Inoue2018_GCal}%
  \BibitemOpen
  \bibfield  {author} {\bibinfo {author} {\bibfnamefont {Y.}~\bibnamefont
  {Inoue}} \emph {et~al.},\ }\href@noop {} {\  (\bibinfo {year} {2018})},\
  \Eprint {http://arxiv.org/abs/1804.08249} {arXiv:1804.08249 [gr-qc]}
  \BibitemShut {NoStop}%
\bibitem [{\citenamefont {Inoue}\ \emph
  {et~al.}(2026{\natexlab{c}})\citenamefont {Inoue}, \citenamefont {Tanabe},\
  and\ \citenamefont {Kumar}}]{inoue2026improvingcalibrationaccuracytorque}%
  \BibitemOpen
  \bibfield  {author} {\bibinfo {author} {\bibfnamefont {Y.}~\bibnamefont
  {Inoue}}, \bibinfo {author} {\bibfnamefont {D.}~\bibnamefont {Tanabe}}, \
  and\ \bibinfo {author} {\bibfnamefont {V.}~\bibnamefont {Kumar}},\
  }\href@noop {} {\  (\bibinfo {year} {2026}{\natexlab{c}})},\ \Eprint
  {http://arxiv.org/abs/2602.19436} {arXiv:2602.19436 [gr-qc]} \BibitemShut
  {NoStop}%
\bibitem [{\citenamefont {Hall}\ \emph {et~al.}(2018)\citenamefont {Hall} \emph
  {et~al.}}]{Hall2017CalibrationRequirements}%
  \BibitemOpen
  \bibfield  {author} {\bibinfo {author} {\bibfnamefont {E.~D.}\ \bibnamefont
  {Hall}} \emph {et~al.},\ }\href {\doibase 10.1088/1361-6382/ab368c}
  {\bibfield  {journal} {\bibinfo  {journal} {Class. Quantum Grav.}\ }\textbf
  {\bibinfo {volume} {35}},\ \bibinfo {pages} {065003} (\bibinfo {year}
  {2018})}\BibitemShut {NoStop}%
\bibitem [{\citenamefont {Cahillane}\ \emph {et~al.}(2017)\citenamefont
  {Cahillane} \emph {et~al.}}]{Cahillane2017_Calibration}%
  \BibitemOpen
  \bibfield  {author} {\bibinfo {author} {\bibfnamefont {C.}~\bibnamefont
  {Cahillane}} \emph {et~al.},\ }\href {\doibase 10.1103/PhysRevD.96.102001}
  {\bibfield  {journal} {\bibinfo  {journal} {Phys. Rev. D}\ }\textbf {\bibinfo
  {volume} {96}},\ \bibinfo {pages} {102001} (\bibinfo {year}
  {2017})}\BibitemShut {NoStop}%
\bibitem [{\citenamefont {Borka}\ \emph {et~al.}(2013)\citenamefont {Borka}
  \emph {et~al.}}]{Borka2013}%
  \BibitemOpen
  \bibfield  {author} {\bibinfo {author} {\bibfnamefont {D.}~\bibnamefont
  {Borka}} \emph {et~al.},\ }\href {\doibase 10.1088/1475-7516/2013/11/050}
  {\bibfield  {journal} {\bibinfo  {journal} {JCAP}\ }\textbf {\bibinfo
  {volume} {11}},\ \bibinfo {pages} {050} (\bibinfo {year} {2013})},\ \Eprint
  {http://arxiv.org/abs/1311.1404} {arXiv:1311.1404} \BibitemShut {NoStop}%
\bibitem [{\citenamefont {Psaltis}\ \emph {et~al.}(2016)\citenamefont {Psaltis}
  \emph {et~al.}}]{Psaltis2016}%
  \BibitemOpen
  \bibfield  {author} {\bibinfo {author} {\bibfnamefont {D.}~\bibnamefont
  {Psaltis}} \emph {et~al.},\ }\href {\doibase 10.3847/0004-637X/818/2/121}
  {\bibfield  {journal} {\bibinfo  {journal} {Astrophys. J.}\ }\textbf
  {\bibinfo {volume} {818}},\ \bibinfo {pages} {121} (\bibinfo {year}
  {2016})},\ \Eprint {http://arxiv.org/abs/1510.00394} {arXiv:1510.00394}
  \BibitemShut {NoStop}%
\bibitem [{\citenamefont {Boehle}\ \emph {et~al.}(2016)\citenamefont {Boehle}
  \emph {et~al.}}]{Boehle2016}%
  \BibitemOpen
  \bibfield  {author} {\bibinfo {author} {\bibfnamefont {A.}~\bibnamefont
  {Boehle}} \emph {et~al.},\ }\href {\doibase 10.3847/0004-637X/830/1/17}
  {\bibfield  {journal} {\bibinfo  {journal} {Astrophys. J.}\ }\textbf
  {\bibinfo {volume} {830}},\ \bibinfo {pages} {17} (\bibinfo {year} {2016})},\
  \Eprint {http://arxiv.org/abs/1607.05726} {arXiv:1607.05726} \BibitemShut
  {NoStop}%
\bibitem [{\citenamefont {Gillessen}\ \emph {et~al.}(2017)\citenamefont
  {Gillessen} \emph {et~al.}}]{Gillessen2017}%
  \BibitemOpen
  \bibfield  {author} {\bibinfo {author} {\bibfnamefont {S.}~\bibnamefont
  {Gillessen}} \emph {et~al.},\ }\href {\doibase 10.3847/1538-4357/aa5c41}
  {\bibfield  {journal} {\bibinfo  {journal} {Astrophys. J.}\ }\textbf
  {\bibinfo {volume} {837}},\ \bibinfo {pages} {30} (\bibinfo {year} {2017})},\
  \Eprint {http://arxiv.org/abs/1611.09144} {arXiv:1611.09144} \BibitemShut
  {NoStop}%
\bibitem [{\citenamefont {Fischbach}\ and\ \citenamefont
  {Talmadge}(1992)}]{Fischbach1992}%
  \BibitemOpen
  \bibfield  {author} {\bibinfo {author} {\bibfnamefont {E.}~\bibnamefont
  {Fischbach}}\ and\ \bibinfo {author} {\bibfnamefont {C.}~\bibnamefont
  {Talmadge}},\ }\href {\doibase 10.1038/356207a0} {\bibfield  {journal}
  {\bibinfo  {journal} {Nature}\ }\textbf {\bibinfo {volume} {356}},\ \bibinfo
  {pages} {207} (\bibinfo {year} {1992})}\BibitemShut {NoStop}%
\bibitem [{\citenamefont {Matone}\ \emph {et~al.}(2007)\citenamefont {Matone}
  \emph {et~al.}}]{Matone2007_CQG}%
  \BibitemOpen
  \bibfield  {author} {\bibinfo {author} {\bibfnamefont {L.}~\bibnamefont
  {Matone}} \emph {et~al.},\ }\href {\doibase 10.1088/0264-9381/24/8/013}
  {\bibfield  {journal} {\bibinfo  {journal} {Class. Quantum Grav.}\ }\textbf
  {\bibinfo {volume} {24}},\ \bibinfo {pages} {2217} (\bibinfo {year}
  {2007})}\BibitemShut {NoStop}%
\bibitem [{\citenamefont {Tanabe}\ \emph {et~al.}(2025)\citenamefont {Tanabe}
  \emph {et~al.}}]{Tanabe2025_CHRONOS_Intensity}%
  \BibitemOpen
  \bibfield  {author} {\bibinfo {author} {\bibfnamefont {D.}~\bibnamefont
  {Tanabe}} \emph {et~al.},\ }\href@noop {} {\  (\bibinfo {year} {2025})},\
  \Eprint {http://arxiv.org/abs/2510.24779} {arXiv:2510.24779 [gr-qc]}
  \BibitemShut {NoStop}%
\bibitem [{\citenamefont {Grosswald}(1978)}]{Grosswald1978}%
  \BibitemOpen
  \bibfield  {author} {\bibinfo {author} {\bibfnamefont {E.}~\bibnamefont
  {Grosswald}},\ }\href@noop {} {\emph {\bibinfo {title} {Bessel
  Polynomials}}}\ (\bibinfo  {publisher} {Springer},\ \bibinfo {year}
  {1978})\BibitemShut {NoStop}%
\end{thebibliography}%
\end{document}